\documentclass[aps,prb,twocolumn,showpacs,preprintnumbers,amsmath,amssymb]{revtex4-1}  %
\usepackage{graphicx}
\usepackage{subfigure}
\usepackage[usenames]{color}
\usepackage[usenames,dvipsnames]{xcolor}
\usepackage{braket}

\emergencystretch=\maxdimen
\begin{document}

\title{The fine structure of the entanglement entropy in the classical XY model}
\author{Li-Ping Yang$^1$}\email[]{liping2012@cqu.edu.cn}
\author{Yuzhi Liu$^{2}$}
\author{Haiyuan Zou$^3$}
\author{Z. Y. Xie$^{4}$}
\author{Y. Meurice$^5$}

\affiliation{$^1$ Department of Physics,Chongqing University, Chongqing 401331, China}
\affiliation{$^2$ Department of Physics, University of Colorado, Boulder, Colorado 80309, USA}
\affiliation{$^3$Department of Physics and Astronomy, University of Pittsburgh, Pittsburgh, PA, USA}
\affiliation{$^4$ Institute of Physics, Chinese Academy of Sciences, P.O. Box 603, Beijing 100190, China}
\affiliation{$^5$ Department of Physics and Astronomy, The University of Iowa, Iowa City, Iowa 52242, USA }

\def\lt{\lambda ^t}
\def\note{note}
\def\beq{\begin{equation}}
\def\enq{\end{equation}}
\def\hata{\hat{\alpha}}
\def\hatx{\hat{x}}
\def\hatt{\hat{t}}
\def\ms{\text{-}s}
\def\mt{\text{-}t}
\def\Tr{{\rm Tr}\,}
\date{\today}

\def\EE{entanglement entropy }
\def\TE{thermal entropy }

\begin{abstract}
We compare two calculations of the particle density in the superfluid phase of the classical XY model with a chemical potential $\mu$  in 1+1 dimensions.
The first relies on exact blocking formulas from
the Tensor Renormalization Group (TRG) formulation of the transfer matrix. The second is a worm algorithm. We show that the particle number distributions obtained with the two methods agree well.  We use the TRG method to calculate the thermal entropy and
the entanglement entropy. We describe the particle density, the two entropies  and the topology of the world lines as we increase $\mu$ to go across the superfluid phase between the first two Mott insulating phases.
For a sufficiently large temporal size, this process reveals an interesting fine structure: the average particle number and the winding number of most of the world lines in the Euclidean time direction increase by one unit at a time.  At each step, the thermal entropy develops a peak and the entanglement entropy increases until we reach half-filling and then decreases in a way that approximately mirror the ascent. This suggests an approximate fermionic picture.

\end{abstract}

\pacs{05.10.Cc,05.50.+q,11.10.Hi,64.60.De,75.10.Hk }
\maketitle
\section{Introduction}
The classical XY model in one space and one Euclidean time (1+1) dimensions plays an important role in the field theoretical approach of condensed matter phenomena and is prominently featured in standard textbooks\cite{kadanoff2000statistical,sachdev2011quantum,chaikin2000principles,herbut2007modern}. This model provides the simplest example of a Berezinski-Kosterlitz-Thouless transition \cite{berezinski,1973kt} and it may be used as an effective theory for the Bose-Hubbard model \cite{PhysRevB.40.546}. More generally, its associated quantum Hamiltonian of Abelian rotors appears in many different contexts such as the formulation of Abelian lattice gauge theories \cite{1987gauge}, Josephson junctions arrays \cite{PhysRevB.61.11289} and cold atom simulators \cite{PhysRevA.90.063603,Bazavov:2015kka}.

When a chemical potential $\mu$ is introduced, the model displays a rich phase diagram depicted in Ref. \onlinecite{PhysRevA.90.063603}.
For sufficiently small $\beta$, the inverse temperature of the classical model, if we increase $\mu$,
we go from a Mott insulating (MI) phase where the average particle number $\rho$ remains zero until it reaches a superfluid (SF) phase where  $\rho$
starts increasing with $\mu$. This proceeds until  $\rho$ reaches one per site and we enter in a new MI phase where it stabilizes at this value. For $\beta$ small enough, this alternation of MI and SF phases repeats several times. In suitable coordinates \cite{PhysRevA.90.063603}, the phase diagram is similar to what is found for the one-dimensional quantum Bose-Hubbard model \cite{PhysRevB.46.9051,PhysRevB.58.R14741}.

\begin{figure}[hh]
 \includegraphics[width=2.5in]{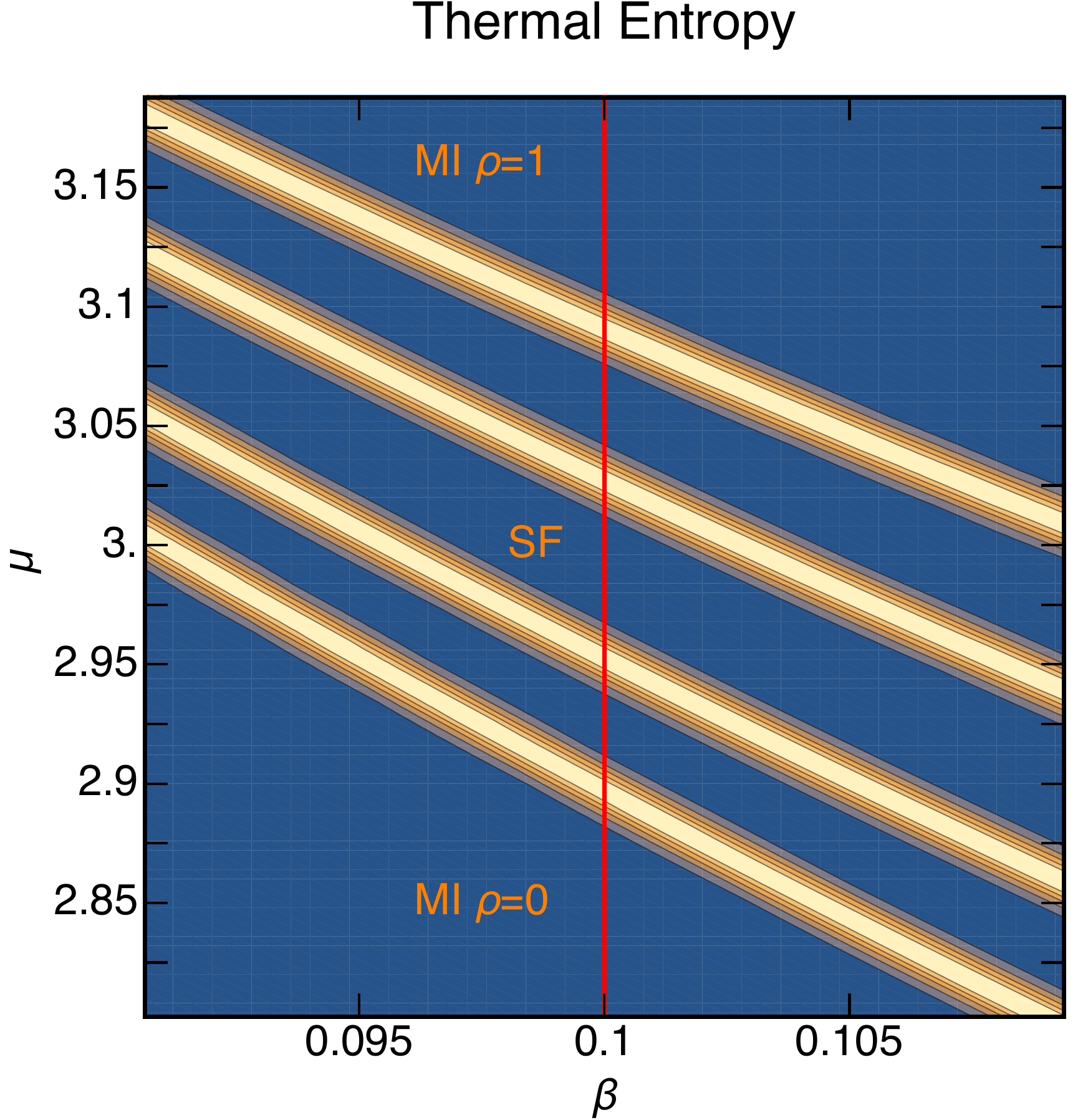}
 \caption{\label{fig:phased2s}  (Color online).  Intensity plot for the thermal entropy of the classical XY model  on a $4\times128$ lattice in the $\beta$-$\mu$ plane. The dark (blue) regions are close to zero and the light (yellow ochre) regions peak near $\ln 2$. The MI phase
 with $\rho$=0 is below the lowest light band, the MI phase with $\rho$=1 is above the highest light band and there is a single SF phase in between the two MI phases.}
\end{figure}

A simple way to locate approximately the SF phase consists in calculating the thermal entropy, defined precisely in Eq. (\ref{Eq:Tentropy}), on a lattice with a large enough temporal size. An example is shown in Fig. \ref{fig:phased2s}. The central line at $\beta$=0.1, where we do calculations hereafter, covers the first SF phase  and the MI phases with $\rho=0$ and 1.
As we increase $\mu$ the thermal entropy goes through a sequence of peaks culminating around  $\ln 2$, signaling level crossings that will be
explained below. The number of peaks equals the number of sites in the spatial direction. This is clearly a finite size feature, while the notion of phase used above should be understood in the limit of an infinite number of sites.

In this article, we describe microscopically the rich sequence of changes which occurs
as we move across the SF phase along a line of constant $\beta$ as described above and illustrated in Fig. \ref{fig:phased2s}. The figure can be embedded in the phase diagrams of Ref. \onlinecite{PhysRevA.90.063603}:  the lower and upper light lines ending at the tips of the
the two MI phases. The fine structure that we report here is first studied for a small spatial size of four sites and then for larger sizes. At infinite spatial size,
there are three phases in Fig. \ref{fig:phased2s}, the SF phase approximately covering the four light bands with larger values of the thermal entropy and the three darker regions in between. We also report about the numerical methods that we developed in this process.

The specific classical XY model used in this article is a planar version of the Ising model with a $O(2)$ symmetry, on a $L_x\times L_t$ lattice. The notations used later to characterize the model and the basic numerical methods are provided in Sec. \ref{sec:modelcalc}. Most of the calculations done in this article will rely  on the tensor renormalization group (TRG) method\cite{PhysRevB.86.045139} which
can be used to write exact blocking formulas for the model considered\cite{YMPRB13,PhysRevD.88.056005,PhysRevE.89.013308} . In Sec.  \ref{subsec:trg}, we remind how the method can be applied to the calculation of the transfer matrix \cite{PhysRevA.90.063603}.

It should be noticed that the presence of $\mu$
causes the action to be complex. This sign problem prevents the use of Monte Carlo simulations when $\mu$ is too large to rely on reweighing methods.
However, using a Fourier expansion of the Boltzmann weights \cite{RevModPhys.52.453}, it is possible to find a formulation of the partition function in terms of
world lines with a positive weight as long as $\mu$ is real. This positivity allows statistical sampling \cite{Banerjee:2010kc} using a classical version of the worm algorithm \cite{PhysRevLett.87.160601}.
The computational methods used for the worm algorithm are briefly reviewed in Sec. \ref{subsec:worm}.

The TRG method  can be used for arbitrary complex values \cite{signtrg} of $\beta$ and $\mu$. The only sources of error are the truncations of the infinite sum to finite ones as required for the numerical treatment. These truncations take place in the original formulation of the partition function and also at each step of the coarse-graining process.  A first check of the agreement between the TRG and the worm methods is provided in Sec. \ref{sec:histo} where we compare particle number density histograms and show that they agree well.

In the SF phase, the model is gapless in the the limit $L_x\rightarrow \infty$. At finite $L_x$ and small $\beta$, the gap is expected to scale like $1/L_x^2$. This
follows from a non-relativistic dispersion relation and can be justified using degenerate perturbation theory. An invaluable method to study the long range correlations in near gapless situations is to calculate the entanglement entropy. Interestingly, it seems possible to measure the \EE  in
many-body systems implemented in cold atoms \cite{PhysRevLett.93.110501,PhysRevLett.109.020505,zollerEE,zollerTE,2015arXiv150400164B,mazza15,fukuhara}.
This includes studies of rather small $L_x$ systems.
The \EE needs to be distinguished from the thermal entropy \cite{zollerTE}. For this reason, in Sec. \ref{sec:TE} we discuss these two entropies using the transfer matrix formalism.
We use a lattice version of the setup of Calabrese and Cardy \cite{Calabrese:2004eu,Calabrese:2005zw} for 1+1 dimensional nonlinear sigma models.
We consider the case of finite, but often large, $L_t$. A finite $L_t$ introduces a temperature proportional to $1/L_t$, distinct from $1/\beta$ used in the classical formulation, and therefore we have a thermal density matrix.
In this context, the relation between the two entropies is a rather open topic of investigation\cite{Cardy:2014jwa}.
Numerical calculations of the related Renyi entropy of the classical XY model, without chemical potential, were presented in Ref. \onlinecite{PhysRevB.87.195134}.

With the TRG method, we approximate the reduced density matrix by a finite dimensional matrix which can be diagonalized numerically and we do not need to use the replica trick as in Refs. \onlinecite{Calabrese:2004eu,Calabrese:2005zw,PhysRevB.87.195134}.  We then show how to use the TRG method to express the \EE for a bipartition of the system in two
subsystems.
We show that for small $L_t$, the \TE is larger than the entanglement entropy but the thermal entropy becomes larger as $L_t$ is increased.

In Sec. \ref{sec:fine}, we use degenerate perturbation theory to get an approximate idea of the large structure (location of SF phases) and
fine structure (changes of  $\rho$ and \EE across one SF phase) of the phase diagram.
We relate in some approximate way, the eigenvectors of the transfer matrix with particle number $n$ to world-line configurations with a winding number which is also $n$.
We then calculate numerically the average particle number density, the \TE and the \EE for values of $\mu$ spanning the
first SF phase as explained above.

For sufficiently large $L_t$, the results show an interesting fine structure: the particle number and the winding number of (most of) the world lines increase by one unit at a time as we keep increasing $\mu$.  At each step, the thermal entropy develops a peak and the entanglement entropy increases with $\mu$ until $\rho$ reaches half-filling. As we keep increasing $\mu$ beyond this value, the \EE decreases in a way  that approximately mirrors its ascent. This approximate symmetry
can be justified by noticing that if we interchange the occupied links with the unoccupied links in the world lines, we transform a configuration with particle number
$n$ to one with a particle number $L_x-n$.
This approximate particle-hole transformation can be reformulated in the context of
degenerate perturbation theory and suggests that a fermionic description is possible.

Our results are summarized in Sec. \ref{sec:con} where we also briefly discuss work in progress. We suggest ways to reduce the small truncation errors
reported in Sec. \ref{sec:histo} and to interpret the approximate particle-hole symmetry found in Sec.  \ref{sec:fine}.
We also briefly discuss the relationship of our work with Polyakov's loop studies \cite{Hands:2010zp,Hands:2010vw} and with recently proposed cold atom experiments \cite{nateprogress}.
\section{Notations and numerical methods}
\label{sec:modelcalc}
In this section, we introduce the classical XY model with a chemical potential. We describe the two numerical methods used in this article. The first is the TRG which relies on coarse-graining, the second is a worm algorithm which relies on sampling.
This section contains important concepts used in the next sections, such as the representation of the partition function in terms of integers labeling Fourier modes attached to the links, also called bounds,  of the lattice, their use in the transfer matrix formulation and their graphical illustration as world lines. The section also contains technical details about the numerical implementations that are given for completeness.
\subsection{The model}
\label{subsec:model}
We consider the classical XY model, sometimes called the $O(2)$ model, with one space and one Euclidean time direction, and a chemical potential $\mu$. One can interpret $\mu$ as the imaginary part of a constant gauge field in the temporal direction.
The sites of the rectangular lattice are labelled $(x,t)$ and the unit vectors  denoted
$\hat{x}$ and $\hat{t}$. The line segments joining two nearest neighbor sites are called links.
The total number of sites is $L_x\times L_t$. In the following, we are typically interested in the case $L_t>>L_x$.
We assume periodic boundary conditions in space and time.
The partition function reads
        \beq
            Z = \int{\prod_{(x,t)}{\frac{d\theta_{(x,t)}}{2\pi}} {\rm e}^{-S}}
            \label{eq:bessel}
        \enq
with
 \begin{eqnarray}
      S=&-&  \beta_{\hat{t}}\sum\limits_{(x,t)} \cos(\theta_{(x,t+1)} - \theta_{(x,t)}-i\mu)\cr&-&\beta_{\hat{x}}\sum\limits_{(x,t)} \cos(\theta_{(x+1,t)} - \theta_{(x,t)}).
      \end{eqnarray}
In all the numerical calculations done in this article, we consider space-time isotropic couplings $\beta_{\hat{x}}=\beta_{\hat{t}}=\beta$.
However, if we set $\beta_{\hat{x}}$ to zero, the model becomes a collection of decoupled solvable models.  For analytical purposes, when $\beta$ is small, it is sometimes convenient to first consider
the solvable case $\beta_{\hat{x}}$=0 and then restore the isotropic situation $\beta_{\hat{x}} = \beta_{\hat{t}}=\beta$ perturbatively (see Sec. \ref{sec:fine}).

 As explained in Refs. \onlinecite{RevModPhys.52.453,Banerjee:2010kc,PhysRevA.90.063603},
one can use the Fourier expansion of $e^{\beta\cos\theta}$ in terms of modified
Bessel function of the first kind and then integrate out the $\theta_{(x,t)}$ variables.
The Fourier indices associated with the links coming out of the site $(x,t)$ in the space and time directions are denoted
$n_{(x,t),\hat{x}}$ and  $n_{(x,t),\hat{t}}$ respectively. They can be interpreted as currents or particle numbers passing through the links.
The partition function can then be expressed
as a sum of product of Bessel functions:
\begin{eqnarray}\nonumber
Z &=& \sum_{\{n\} }\prod_{(x,t)} I_{n_{(x,t),\hat{x}}}(\beta_{\hat{x}})I_{n_{(x,t),\hat{t}}}(\beta_{\hat{t}}){e^{\mu n_{(x,t),\hat{t}}}}\\
& \times&\delta_{n_{(x-1,t),\hat{x}}+n_{(x,t-1),\hat{t}},n_{(x,t),\hat{x}}+n_{(x,t),\hat{t}}} \ .
\label{eq:Z_worm}
\end{eqnarray}
The Kronecker delta function in Eq.~(\ref{eq:Z_worm}) ensures the local current conservation and the terms in the partition function can be interpreted as current  loops (also called world lines) which can then be statistically sampled~\cite{Banerjee:2010kc}.
For a system with periodic boundary condition in both space and time directions, as considered
in this paper, the world line can wind around in both directions. The winding numbers
are important to understand the superfluid properties of the system~\cite{PhysRevB.36.8343}.
\begin{figure}[h]
\includegraphics[width=0.7\columnwidth]{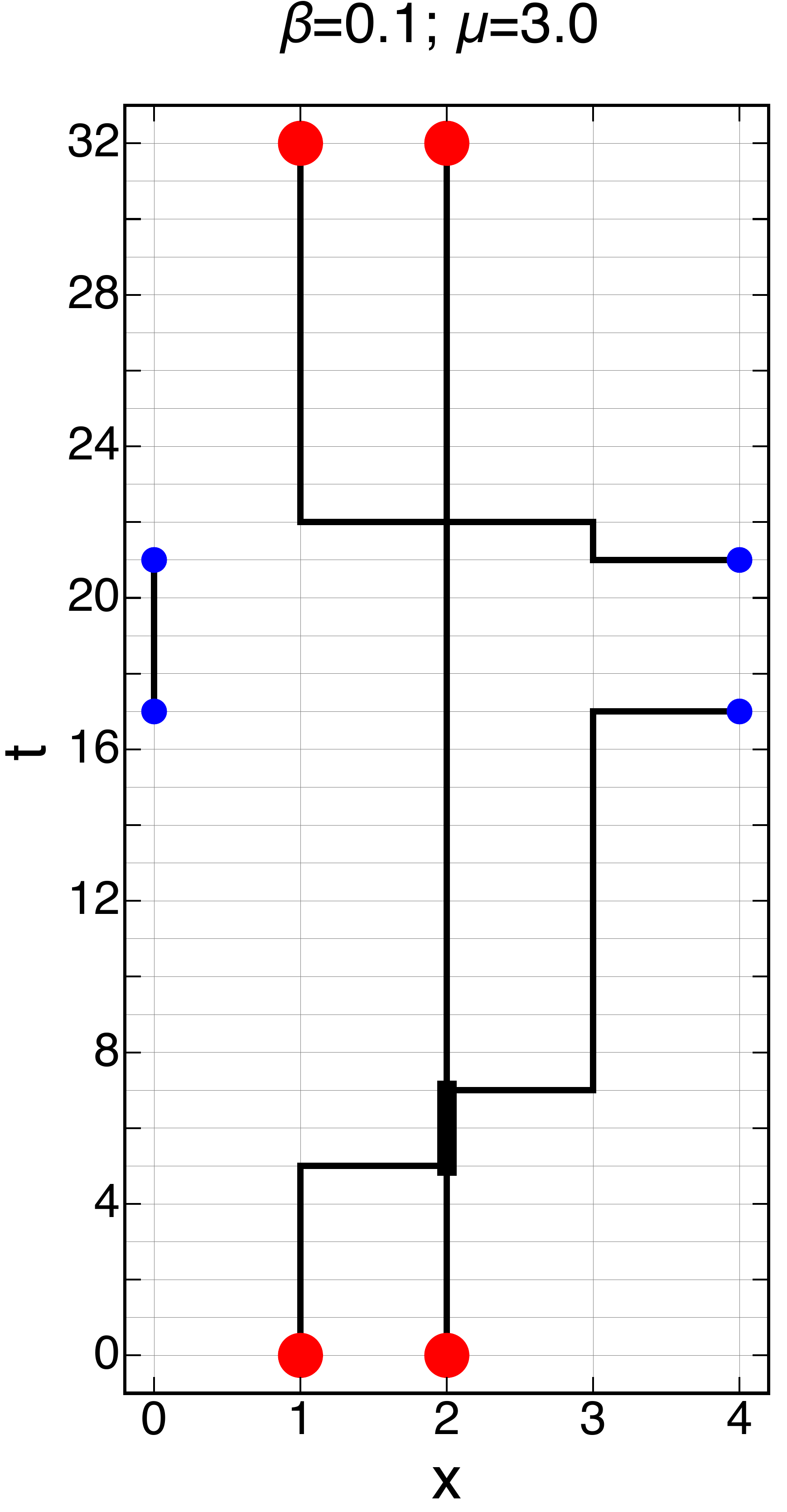}
\caption{\label{fig:typical}  (Color online). Graphical representation of an allowed configuration of $\{n\}$ for a 4 by 32 lattice.
The uncovered links on the grid have $n$=0, the more pronounced dark lines have $|n|$=1 and the wider lines have $n$=2.
The signs of the links with $|n|$=1
lines are discussed in the text.
The large dots on the
horizontal boundaries (red) need to be identified in pairs with the same $x$ coordinates. Similarly, the slightly smaller dots (blue)
on the vertical boundaries have to be identified in pairs with the same $t$ coordinate.}
\end{figure}

An typical allowed configuration for $\beta$=0.1 and $\mu=3$ is shown in Fig. \ref{fig:typical} for $L_x$=4 and $L_t$=32.
Sites at the boundary should be identified as explained in the figure caption.
The uncovered links on the grid have $n$=0, the more pronounced dark lines have $|n|$ =1 and the wider lines have $n$=2.
A discussion regarding the sign convention and the spatial winding number of Fig. \ref{fig:typical} can be found in Appendix \ref{app:a}.

This configuration can be used to visualize a transfer matrix that connects consecutive time slices. For instance, in Fig.  \ref{fig:typical}, the time slice 5 represents a transition between $\ket{1100}$ and $\ket{0200}$ and its relative statistical weight can be obtained from a transfer matrix that will be discussed in Sec. \ref{subsec:trg}.
The configuration was generated using a sampling method designed in Ref. \onlinecite{Banerjee:2010kc} and briefly discussed in  Sec. \ref{subsec:worm}.
 \subsection{TRG approach of the transfer matrix}
 \label{subsec:trg}

\begin{figure}[!h]
\vskip-2cm
 \includegraphics[scale=0.3]{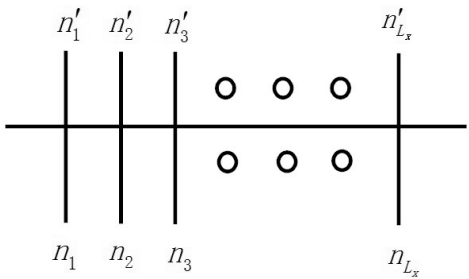}
 \vskip-2cm
 \caption{\label{fig:transfer} Graphical representation of $\mathbb{T}_{(n_1,n_2,\dots, n_{L_x})(n_1',n_2',\dots,n_{L_x}')}$}
\end{figure}

As explained in Ref.   \onlinecite{PhysRevA.90.063603}, the partition function can be expressed in terms  of a transfer matrix:\beq
Z=\Tr\mathbb{T}^{L_t}  \ .
\label{eq:tm}
\enq
The matrix elements of $\mathbb{T}$ can be expressed as a product of tensors associated with the sites of a time slice (fixed $t$) and traced over the space indices.
To make the equations easier to read, we  use the notations $n_x$ for the time indices in the past ($n_{(x,t-1),\hat{t}}$),  the primed symbol $n'_x$ for the time indices in the future ($n_{(x,t),\hat{t}}$) and $\tilde{n}_x$ for the space indices ($n_{(x,t),\hat{x}}$). The matrix elements of $\mathbb{T}$  have the explicit form
\begin{eqnarray}
&\ &\mathbb{T}_{(n_1,n_2,\dots, n_{{L_x}})(n_1',n_2',\dots, n_{L_x}')}=\cr &\ &\cr&\ &\sum_{\tilde{n}_{1}\tilde{n}_{2}\dots \tilde{n}_{L_x}} T^{(1,t)}_{\tilde{n}_{L_x}\tilde{n}_{1}n_1 n_1'}T^{(2,t)}_{\tilde{n}_{1}\tilde{n}_{2}n_2n_2'\dots }\cr &\ &\dots  T^{(L_x,t)}_{\tilde{n}_{L_{x-1}}\tilde{n}_{L_x}n_{L_x}n_{L_x}' },
\label{eq:tmspace}
\end{eqnarray}
with
        \begin{eqnarray}
        T^{(x,t)}_{\tilde{n}_{x-1}\tilde{n}_x n_x n_x'} &=&  \sqrt{I_{n_{x}}(\beta_{\hat{t}})I_{n'_x}(\beta_{\hat{t}})\exp(\mu(n_x+n_x'))}\nonumber \\
            & \ &\sqrt{I_{\tilde{n}_{x-1}}(\beta_{\hat{x}})I_{\tilde{n}_x}(\beta_{\hat{x}})} \delta_{\tilde{n}_{x-1}+n_x,\tilde{n}_x+n_x'} \ .
             \label{eq:tensor}
        \end{eqnarray}
In Eq. (\ref{eq:tm}), the trace is over the temporal indices,
represented graphically by vertical lines in Fig. \ref{fig:transfer}, while the spatial indices (horizontal links in  Fig. \ref{fig:transfer}) are summed over as described in Eq. (\ref{eq:tmspace}).
The Kronecker delta function in Eq. (\ref{eq:tensor}) is the same as in Eq. (\ref{eq:Z_worm}) and reflects the existence of a conserved current as discussed above, that we will call ``particle number" in the following. For periodic (spatial trace) or
open (zero spatial indices at both ends), the local conservation law implies that the transfer matrix elements are zero unless the sum of the two sets of indices (respectively denoted $n$ and $n'$ in Eq. (\ref{eq:tmspace})) are equal. In other words, the
transfer matrix is block diagonal in each particle number sector
\beq
n=\sum_{x=1}^{L_x }n_{x}=\sum_{x=1}^{L_x }n'_{x}\ .
\enq

When the chemical potential $\mu$ is zero, there is a charge conjugation symmetry which allows us to change the sign of all the $n$'s  in all the temporal sums without affecting the final results.
Given the rapid decay of the modified Bessel function when the index $n$ increases, good approximations can be obtained
by replacing the infinite sums by sums restricted from $-n_{max}$ to $n_{max}$. When the chemical potential is nonzero, it is more efficient to shift
the range in the same direction as the sign of the chemical potential.
In general, we call $D_s$ the number of states kept after truncation. In the following, we will use a coarse-graining procedure for the transfer matrix where we repeatedly apply a truncation at the same value $D_s$.

\begin{figure}[h]
 \includegraphics[width=0.75\columnwidth]{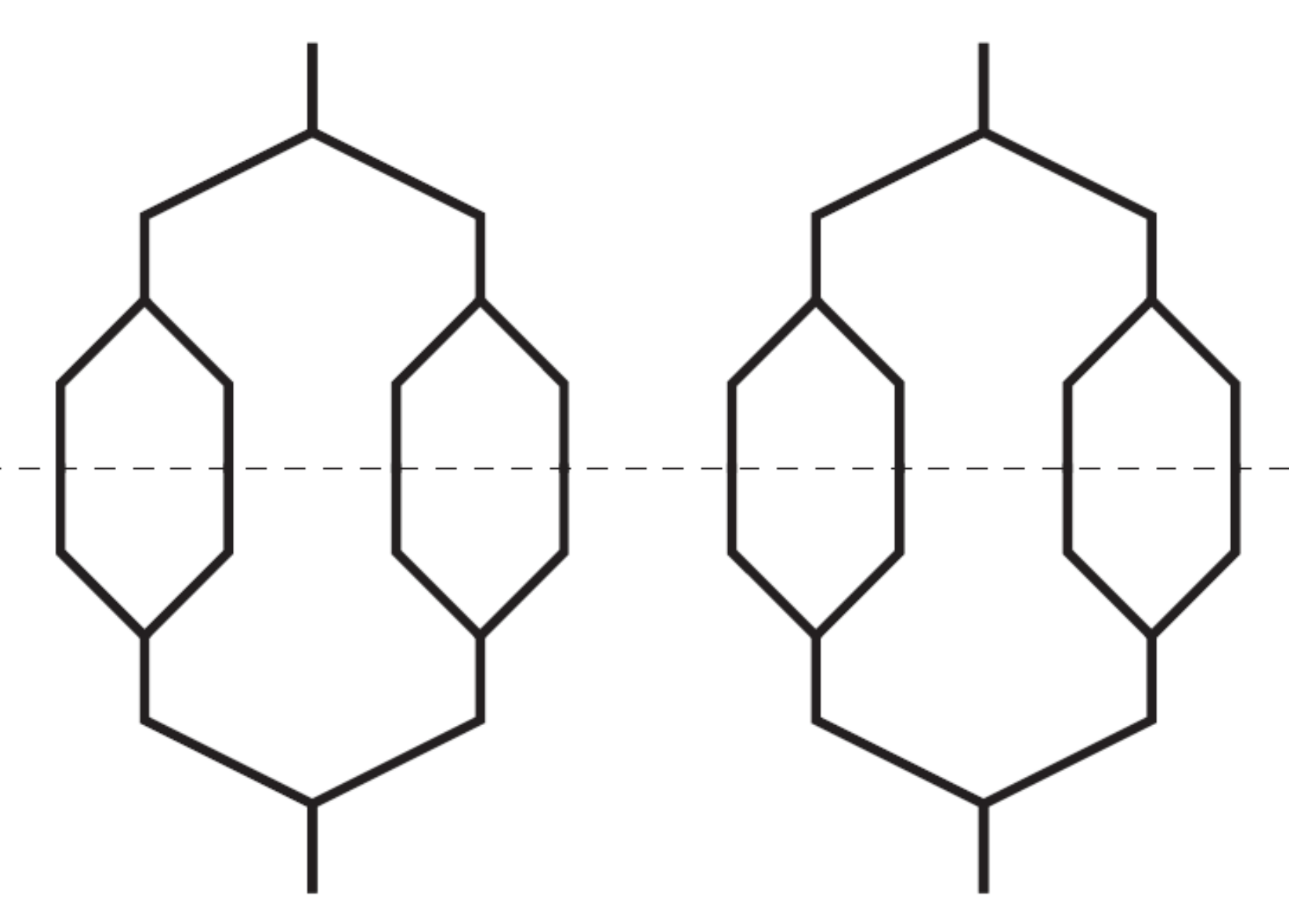}
\caption{\label{fig:blocked} Graphical representation of the coarse graining truncation of the transfer matrix described in the text.}
\end{figure}

For numerical purposes, we reduce the size of the transfer matrix by using a blocking procedure
\cite{PhysRevB.86.045139,PhysRevD.88.056005}.
It consists in iteratively replacing blocks of size two in the transfer matrix
by a single site using a higher order singular value decomposition. During the truncation, for a given pair of sites, we replace the direct product $D_s^2\times D_s^2$ matrix by a $D_s\times D_s$
matrix. The process is illustrated in Fig. \ref{fig:blocked}. The Y-shaped parts in the bottom part of the figure and their mirror versions in the top part schematically represent this truncation.
The full technical details can be found in Refs. \onlinecite{PhysRevB.86.045139,PhysRevD.88.056005}.

\subsection{The worm algorithm}
\label{subsec:worm}

Because of the conservation law in Eq. (\ref{eq:tensor}), the terms of the partition function can be interpreted as current loops which can be statistically sampled
\cite{Banerjee:2010kc}. The sampling procedure for one complete worm algorithm update goes as follows. We pick randomly a site on the lattice, then
pick randomly a neighboring direction in the (positive or negative) spatial or time direction, change the current $n$ to $n\pm1$ (depending on the sign of the change)  and move to the neighboring site if the above update is accepted (the details of the accept-reject procedure are given in the Appendix A of Ref. \onlinecite{Banerjee:2010kc}).
We repeat the above procedure until we come back to the original random site. Because of the conservation law, a particle number can be attributed to a configuration. It can be calculated by summing the temporal $n$'s between any two time slices.  The particle number distribution, can be calculated by generating a large number of configurations according to the above procedure.

An easily controllable source of error is the limited statistics.
There is also an unavoidable truncation error. Following Ref. \onlinecite{Banerjee:2010kc},
$|n|$ is constrained to be not larger than 20 in Eq.~(\ref{eq:Z_worm}).
However, given that $I_{20}(0.1)\exp(3\times 20)/I_0(0.1)\simeq 4 \times 10^{-19}$, this is negligible for our calculations. By construction, there is a single worm and the algorithm cannot generate disconnected loops. This is not believed to be a significant source of error. Comparisons at small volume where the truncation errors are controllable \cite{Meurice:2014tca} indicate that the worm algorithm is statistically exact.

\section{Particle density calculations}
\label{sec:histo}

In this section, we show how to calculate the average particle density using the TRG formulation. We then apply the method numerically and compare the results with the ones obtained with the worm algorithm.
The particle number conservation can also be exploited in the TRG approach.
For the initial one-site tensor $T^{(x,t)}$, we have
\begin{equation}
{T'}^{(x,t)} \equiv \partial T^{(x,t)}/\partial\mu = \frac{1}{2}(n_x + n_{x}')T^{(x,t)}\ .
\end{equation}
Here, $n_x$ and $n_x'$  are the particle number associated with the time indices of the original tensor $T^{(x,t)}$ with the tensor indices omitted.
Consequently, we can associate a particle number $n(i)$ with each eigenvalue $\lambda_i$ of the transfer matrix $\mathbb{T}$.
The average particle number density is an extensive quantity defined as
\begin{equation}
\rho=\frac{1}{L_x L_t}\frac{\partial \ln Z}{\partial \mu}. \
\end{equation}
From the expression of $Z$ in terms of the transfer matrix and the cyclicity of the trace, we have
\begin{equation}
\partial Z/\partial\mu= L_t {\rm Tr}(\mathbb{T}'\mathbb{T}^{L_t-1}),
\end{equation}
where $\mathbb{T}' =\partial \mathbb{T}/\partial \mu$ can be calculated by using the chain rule in Eq. (\ref{eq:tmspace}).
This can be achieved iteratively by defining an ``impurity" tensor initialized with the derivative of the initial tensor and then
blocked and symmetrized with the original ``pure" tensor. This guarantees the recursive replacement of $T^{(x,t)}$ by ${T'}^{(x,t)}$ in the transfer matrix as
prescribed by the chain rule.
This procedure can be shortcut if we know the particle number $n(i)$ associated with each eigenvalue $\lambda_i$ as discussed above.
We can then write
\begin{equation}
\frac{1}{L_t}\frac{\partial \ln Z}{\partial\mu}=\frac{\sum_i\lambda_i^{L_t}n(i)}{\sum_i \lambda_i ^{L_t}}.
\end{equation}
In practice, finding the particle number associated with the eigenvalues is not completely straightforward. It requires to keep track of
the particle number in the projected basis or to write the blocking algorithm sector by sector.
The enforcement of the conservation law in the coarse-graining process makes the numerical calculation more stable. The tensor elements violating the conservation law are exactly zero. If we   only handle
non-zero elements, we can reach relatively larger values of the dimension $D_s$ in the truncation
procedure and get more accurate results. These improvements have been used in the numerical calculations done below.

Knowing the $n(i)$ associated with $\lambda_i$ also allows us to define a probability $P(n)$ for the particle number $n$:
\begin{equation}
P(n)=\sum_{i:n(i)=n}\lambda_i ^{L_t}/\sum_{i}\lambda_i^{L_t}.
\end{equation}
These probabilities can also be calculated directly from histograms obtained with the worm algorithm.
Both methods can be used to calculate and compare the average particle number density using
\begin{equation}
\rho=\frac{1}{L_x}\sum_n nP(n)\ .
\end{equation}

\begin{figure}
\advance\leftskip 0.1cm
\vskip-3.7cm
\includegraphics[width=3.8in]{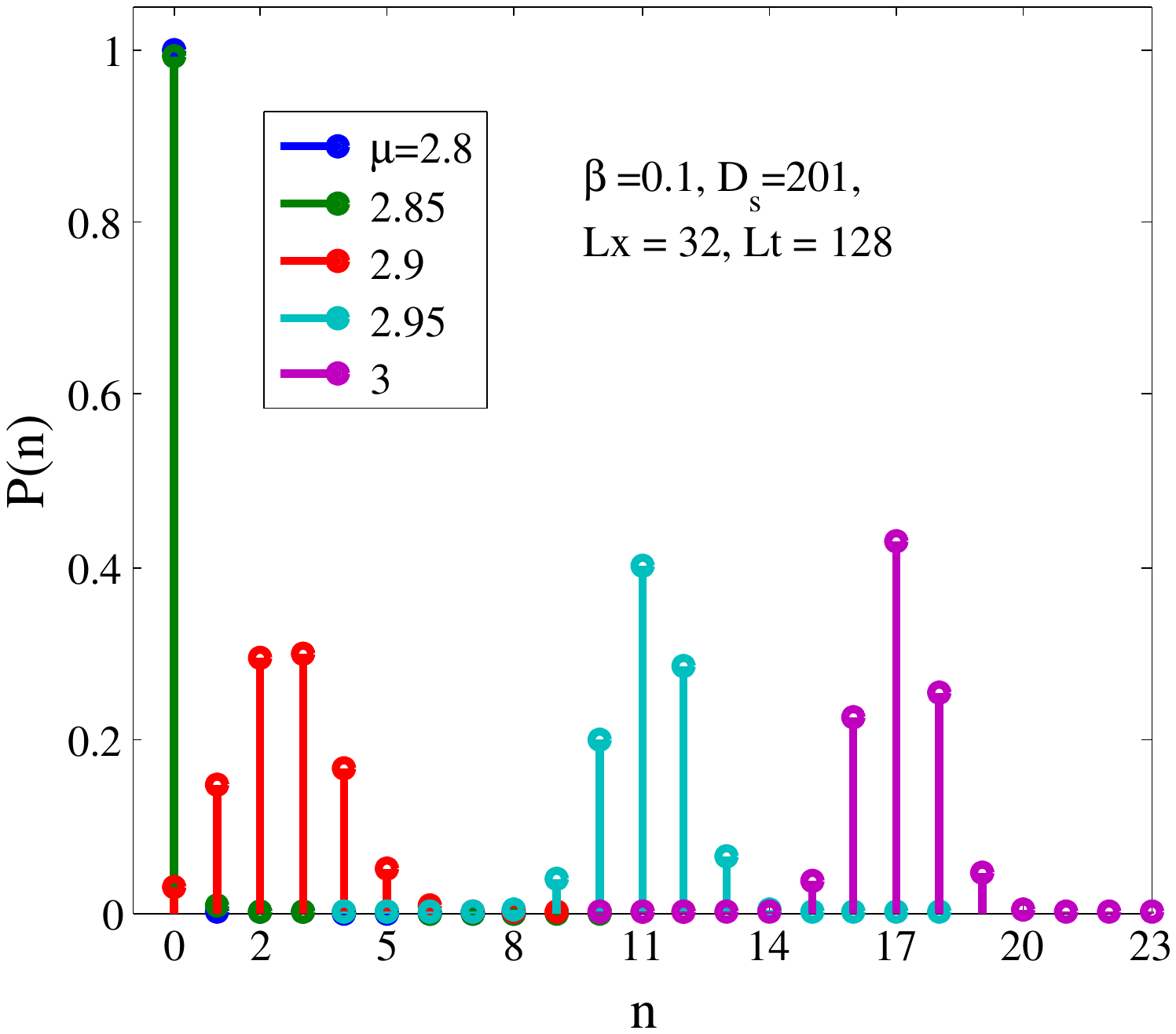}
\vskip-3.5cm
\caption{\label{fig:diffmuNp} (Color online). The particle number distribution $P(n)$ with $\mu$ taking different values at the boundary between the MI and SF phases. For $\mu=2.80$ and 2.85, we only have $n=0$. For  $\mu=2.90$, 2.95 and 3.0, there are three visible groups of bins. With $\mu$ increasing, the distribution shifts to the right with larger most probable particle number. }
\end{figure}

\begin{figure}
\vskip-0.5cm
\includegraphics[width=1.2\columnwidth]{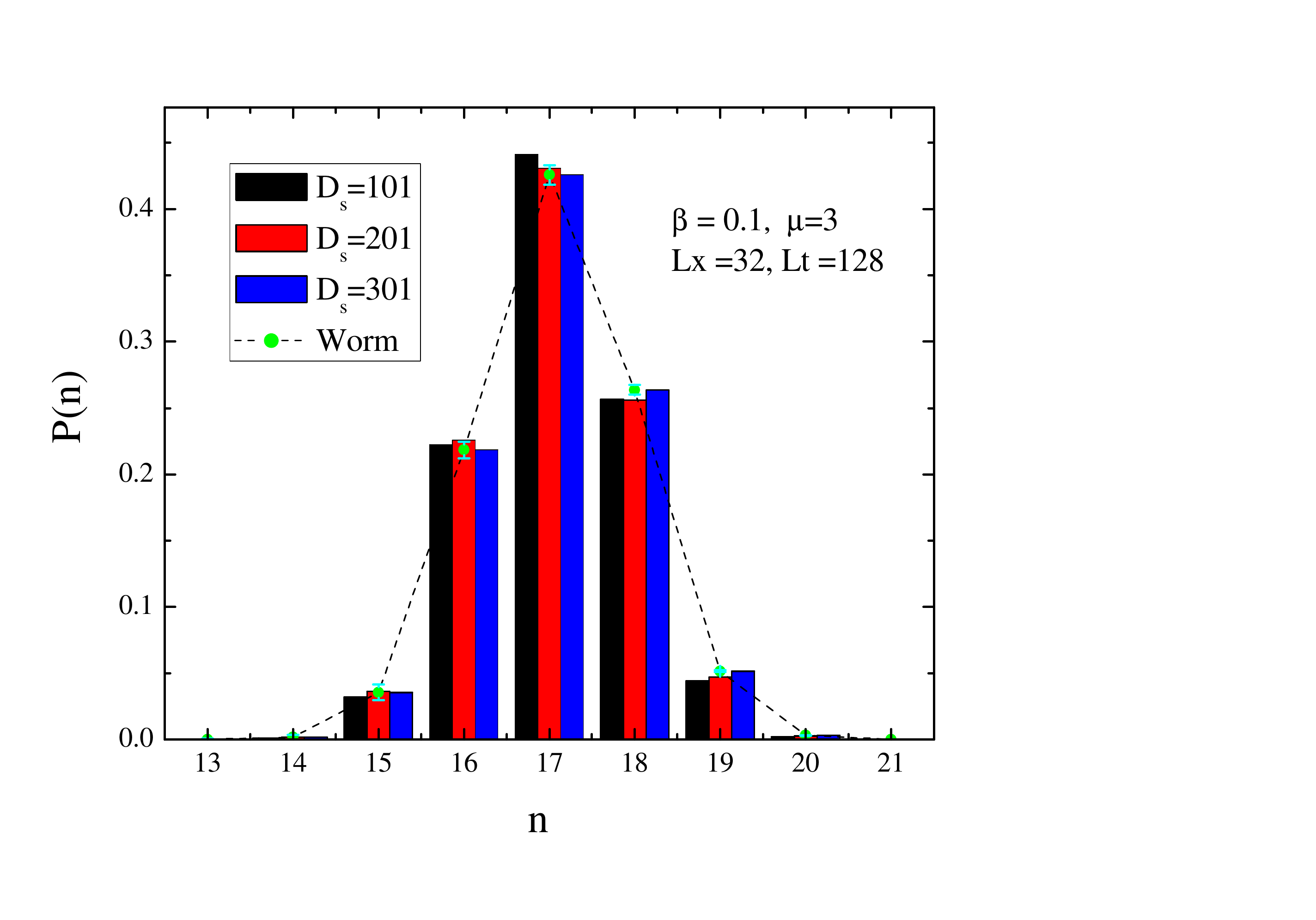}
\vskip-0.5cm
\caption{\label{fig:Np} (Color online). Comparison of the particle number distribution $P(n)$ from the worm algorithm and TRG with different $D_s$. }
\end{figure}

We studied numerically the distribution of $P(n)$ for various  values of $\mu$ spanning a range covering the boundary of the  MI phase and the  SF phase.  The other parameters are kept fixed at
$D_s=201$, $\beta=0.1$, $L_x=32$, and $L_t=128$.
When $\mu=2.8$ and 2.85, the distribution bears one-bin structure with $n=0$, corresponding to the MI phase.
When $\mu=2.9$, 2.95 and 3, the distribution carries more bins, corresponding to the SF phase, which shows that the phase transition
occurs in the range $\mu=(2.85, 2.9)$ as illustrated in Fig. \ref{fig:diffmuNp}.
We then compared the distributions obtained with the TRG and the worm algorithm for $\mu$=3, which is near the middle to the SF phase. The results are shown in Fig. \ref{fig:Np}. To the best of our knowledge, the errors associated with the worm calculations are purely statistical.
The errors made with the TRG are due to the repeated truncation to $D_s$ indices. We have used $D_s$=101, 201 and 301  and the variations are comparable to the worm errors. Overall, the results from $D_s=301$ are very close to the worm calculation. The particle number distribution calculated from the worm histograms  are from averaging over configurations generated with twelve different initial random seeds. With each seed, four million equilibrium configurations were generated. With the spatial dimension increasing, the accuracy becomes worse by keeping the same truncation dimension $D_s$. When the time dimension $L_t$ becomes large enough,
the particle density distribution becomes more consistent for the cases with different truncation dimension, because it is almost centralized in one bin.
\section{Calculation of the thermal and entanglement entropy}
\label{sec:TE}
In this section we explain how to compute the thermal entropy and the entanglement entropy in a consistent way.
We construct the reduced density matrix following Refs. \onlinecite{Calabrese:2004eu,Calabrese:2005zw}.
More specifically, we consider the path integral representation of the thermal density matrix (Eq. (6) in Ref. \onlinecite{Calabrese:2004eu}) for the classical XY model. However, we use the $(n_1,n_2,\cdots,n_{L_x})$ representation and the corresponding transfer matrix introduced in Sec. \ref{subsec:trg} rather than the original spin variables.  In the following, the eigenstates of the transfer matrix will be treated as quantum states.

We consider the system, denoted $AB$, and subdivide it into two parts denoted $A$ and $B$.
We first define the thermal density matrix $\hat{\rho}_{AB}$ for the whole
system
\begin{equation}
\hat{\rho}_{AB}\equiv\mathbb{T}^{L_t}/Z\ .
\end{equation}
We have the usual normalization ${\rm Tr}~\hat{\rho}_{AB}=1$. If the largest eigenvalue of the transfer matrix is non degenerate with an eigenstate denoted $\ket{\Omega}$, we have
the pure state limit
\beq
 \lim_{L_t \rightarrow \infty} \hat{\rho}_{AB} =\ket{\Omega} \bra{\Omega}.
 \enq
In the following, we will work at finite $L_t$ and will deal with the entanglement of thermal states \cite{RevModPhys.82.277}.
In general, the eigenvalue spectrum $\{\rho_{{AB_i}}\}$ of  $\hat{\rho}_{AB}$ can then be used to define the thermal entropy
\begin{equation}
S_T=-\sum_i \rho_{AB_i} \ln(\rho_{AB_i}).\label{Eq:Tentropy}
\end{equation}
The subdivision of $AB$ into $A$ and $B$ refers to a subdivision of the spatial indices.
We define the reduced density matrix $\hat{\rho}_A$ as
\begin{equation}
\hat{\rho}_A\equiv {\rm Tr}_B \hat{\rho}_{AB}.
\end{equation}
We define the \EE of $A$ with respect to $B$ as the von Neumann entropy of this reduced density matrix $\hat{\rho}_A$.
The eigenvalue spectrum $\{\rho_{A_i}\}$ of the reduced density matrix can then be used to calculate the \EE
\begin{equation}
S_E=-\sum_i \rho_{A_i }\ln(\rho_{A_i}).\label{Eq:entropy}
\end{equation}

The computation of the \TE can be performed using the eigenvalues of the transfer matrix $\lambda_i$ discussed in the previous section together with the normalization
$\rho_i=\lambda_i ^{L_t}/\sum_{j}\lambda_j^{L_t}$. Note that if we introduce a temperature $T$ and energy levels by identifying $\lambda_i^{L_t}$ with
$\exp(-E_i/T)$, we recover the standard relation $S_T=\braket{E}/T +\ln Z$. This identification makes clear that $T\propto 1/L_t$. Later, in Eq.(\ref{eq:el}), we will use units where $T=1/L_t$.

\begin{figure}
\includegraphics[width=0.6\columnwidth]{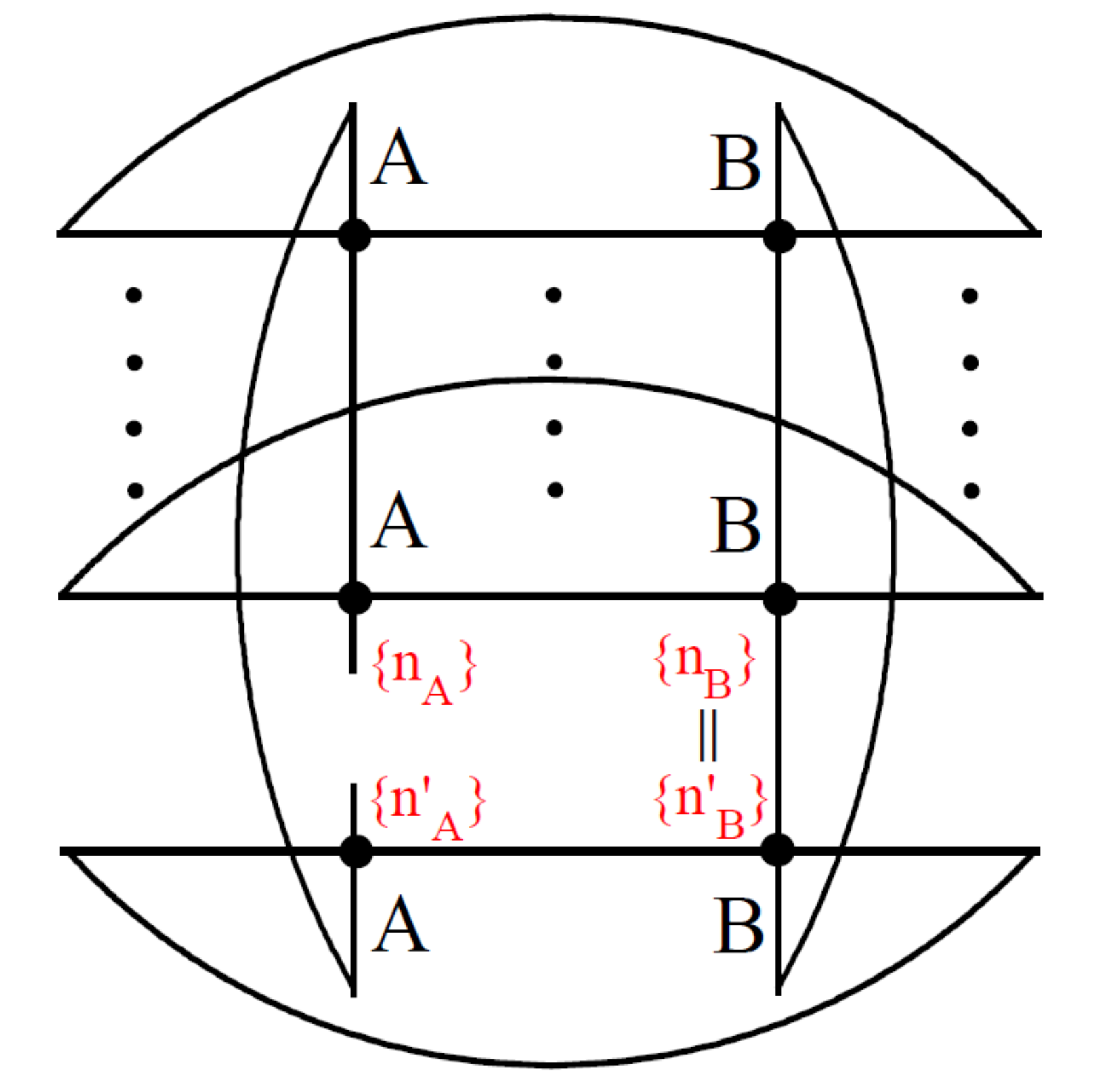}
\caption{\label{fig:EE-demon} (Color online). Illustration of the entanglement entropy calculation. The horizontal lines represent the traces on the space indices. There are $L_t$ of them, the missing ones being represented by dots. The two vertical lines represent the traces over the blocked time indices in $A$ and $B$.}
\end{figure}

For the computation of the \EE, we assume that $A$ has $2^{\ell_A}$ sites and B has $2^{\ell_B}$ sites and then perform the $\ell_A$ and $\ell_B$ blockings  for the subsystems.
The coarse-graining along the spatial direction ends with two sites, one for $A$ and the other for $B$. We can then contract the indices from the two sites in the time direction without further truncation.
Tracing over the space indices liking $A$ and $B$, we obtain the transfer matrix $T_{(\{n_A\},\{n'_A\},\{n_B\},\{n'_B\})}$ for the whole $AB$ system. Taking the $L_t$ power, tracing over $n_B$, and normalizing, we obtain the
reduced density matrix $\hat{\rho}_A(\{n_A\},\{n'_A\})$. This is illustrated in Fig.~\ref{fig:EE-demon}.

In the following, we specialize to  the case of $\ell_A=\ell_B$.  Assembling subsystems of different sizes is not difficult because the space indices are not renormalized in the transfer matrix approach, but this will not be done here.

\begin{figure}
\advance\leftskip 0.45cm
\vskip-4.05cm
\includegraphics[width=4.2in]{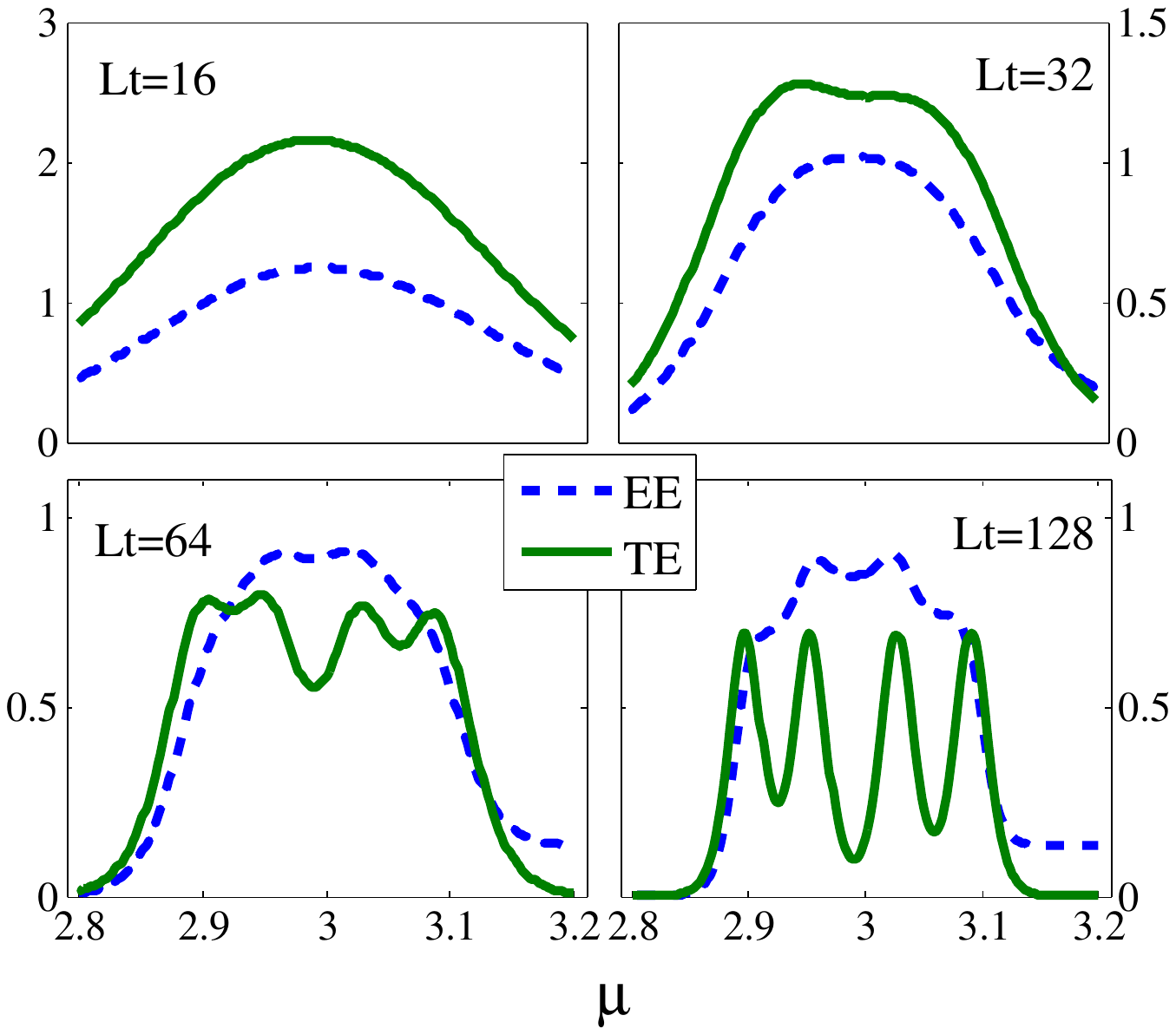}
\vskip-4.8cm
\caption{\label{fig:EETE} (Color online). Entanglement entropy (EE, dash line)  and \TE (TE, solid line) for $\beta=0.1$, $L_x=4$ and $L_t=16$, 32, 64 and 128.
}
\end{figure}

As a first study, we have considered the cases $\beta=0.1$, $L_x=4$ and $L_t=16$, 32, 64 and 128 with $\mu$ below 3.2. Both the \TE and \EE develop a peak over the SF phase.
 We see that for small $L_t$, the \TE is larger than the entanglement entropy, but as we increase $L_t$, the \EE becomes larger than the thermal entropy.
 The results are shown in Fig. \ref{fig:EETE}.
 For both the \TE and the entanglement entropy, a fine structure appears for large $L_t$. This is discussed in the next section.

\section{The fine structure of the SF phase}
\label{sec:fine}

In this section, we first give an idea of the large structure of the phase diagram (where the different SF and MI phases are located) and explain
in more details the fine structure of a single SF phase, more specifically along a line of constant $\beta$ and where $\mu$ is varied in order to interpolate between the
the two MI phases with successive values of $\rho$.

At small $\beta$, we have mostly MI phases with small SF phases in between them.
A good qualitative picture can be obtained by temporarily considering the anisotropic case $\beta_{\hat{x}}=0$ where the problem is
exactly solvable and then restoring $\beta_{\hat{x}}=\beta_{\hat{t}}=\beta$ perturbatively. When $\beta_{\hat{x}}=0,$ we have only onsite interactions and in the large $L_t$ limit, the problem reduces to finding the value $n^\star$ of $n$ which maximizes $I_n(\beta_{\hat{t}}){\rm e}^{\mu n}$ for given $\mu$ and $\beta_{\hat{t}}$. The largest eigenvalue of $\mathbb{T}$ is then $(I_{n^\star}(\beta_{\hat{t}}){\rm e}^{\mu n^\star })^{L_x}$. The quantum picture is that for large $L_t$, the relevant state has all $L_x$ sites in the $n^\star$ state and we
are in the MI phase with $\rho=n^\star$. In summary,
the approximate large structure at small $\beta$ is obtained by increasing $\mu$ from zero and going through the MI phases with $n^\star$= 0, 1, 2, $\dots$ .

The fine structure of the SF phase between the MI phases can be approached by restoring $\beta_{\hat{x}}=\beta_{\hat{t}}$ perturbatively. The SF phases are approximately located near values of $\mu$ where $n^\star$ changes. To be specific, we will consider the example of $\beta=0.1$, $L_x=4$, where the transition occurs near $\mu_c=2.997\cdots $when $\beta_{\hat{x}}=0$.
In this limit, we have 16 degenerate states $\ket{0,0,0,0},\  \ket{1,0,0,0},\ \dots ,\  \ket{1,1,1,1}$ which can be organized in
``bands" with $n=0$ (1 state), $n=1$ (4 states), etc. Below, we call the approximation where the indices inside the kets are only 0 or 1 the ``two-state approximation".

\begin{figure}
\vskip-0.3cm
\advance\leftskip 0.4cm
\includegraphics[width=1\columnwidth]{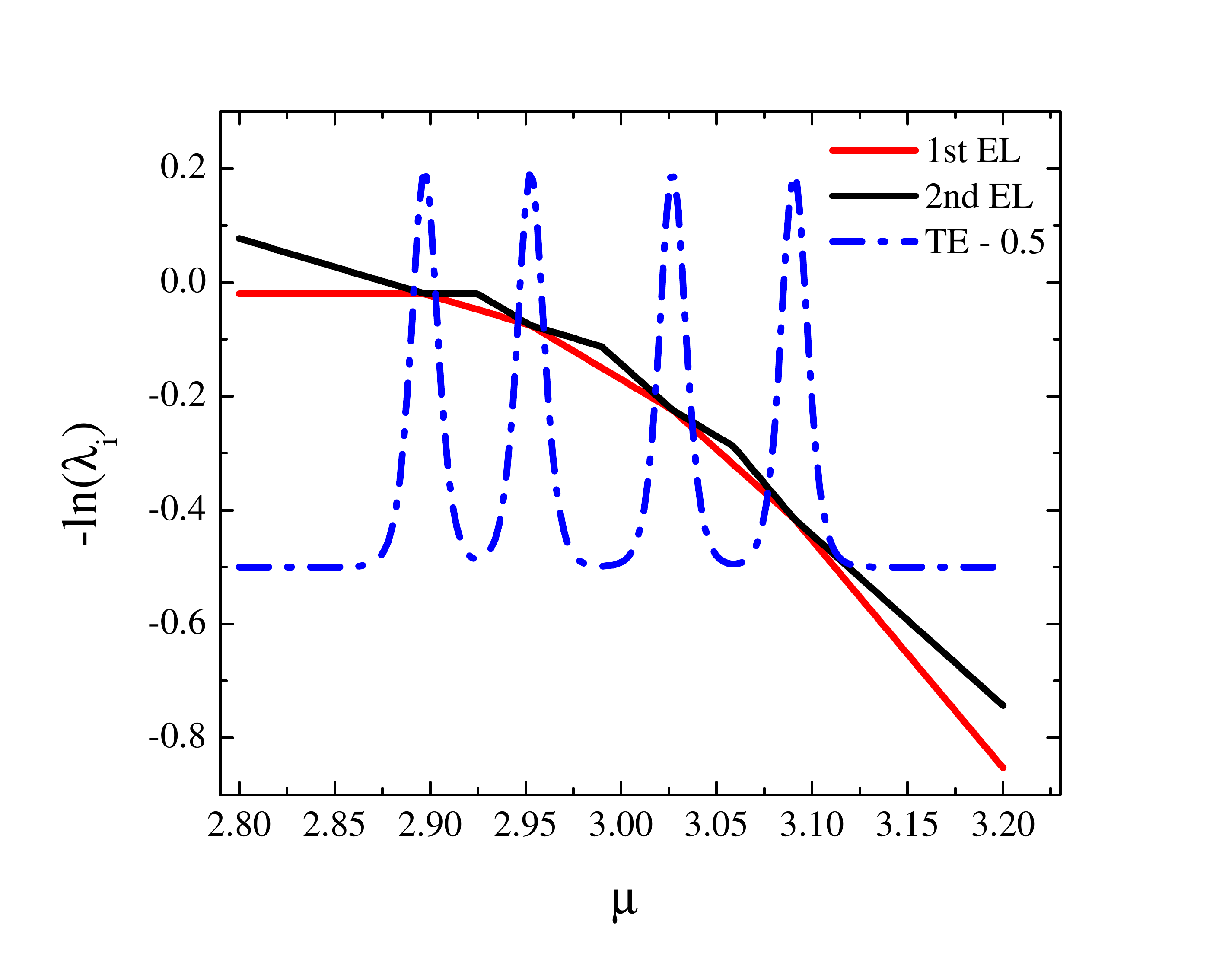}
\caption{\label{fig:EL} (Color online).
The two lowest energy levels (EL) as a function of $\mu$ for $L_x=4,\  L_t=256,\ \beta=0.1 \ {\rm and}\  D_s=101$. As $\mu$ increases, lines of successive
slopes 0, -1, -2, -3, and -4 are  at the lowest level. At each crossing, the thermal entropy jumps.
The values of \TE are shifted vertically by -0.5 to make the figure readable. }
\end{figure}

The effect of $\beta_{\hat{x}}$ is to give these bands a width and lift the degeneracy. The energy levels are defined in terms of the eigenvalues of the transfer matrix as
\begin{equation}
\label{eq:el}
E_i=-\ln(\lambda_i).
\end{equation}
If we plot the energy levels versus $\mu$, we see that we have successive crossings corresponding to
states of increasing $n$. This is illustrated with the two lowest energy levels in Fig. \ref{fig:EL}. Notice the piecewise linear behavior with slopes corresponding to the particle number 0, -1, -2, -3 and -4. As the levels cross the thermal entropy rises to $\ln2$.

We observe that near $\mu=2.90$, the lowest energy state changes from $\ket{0000}$ to a state with
$n=1$
\begin{equation}
\ket{\Omega, n=1}=\frac{1}{2}(\ket{1000 } +\ket{0100}+\ket{0010}+\ket{0001})\ .
\end{equation}
It is easy to calculate the reduced density matrix for $A$ defined as first two sites and $B$ as the last two sites in the limit where
$L_t$ becomes infinite and for values of $\mu$ where  $\ket{\Omega, n=1}$ is the unique ground state
\begin{eqnarray}
\hat{\rho}_A&=&{\rm Tr}_B \ket{\Omega, n=1}\bra{\Omega, n=1}\\ \nonumber
&=&\frac{1}{4}(\ket{10}+\ket{01})(\bra{10}+\bra{01})+\frac{1}{2}\ket{00}\bra{00}\ .
\end{eqnarray}
The eigenvalues of $\hat{\rho}_A$ are 1/2, 1/2, 0, and 0 and the entanglement entropy of this reduced density matrix is $\ln 2$.

A $n=2$ state becomes the ground state near $\mu$=2.95. It is in good approximation a linear superposition of the 6 states with two
0's and two 1's. The two states $\ket{1010}$ and $\ket{0101}$ have a slightly larger coefficient suggesting weak repulsive interactions. In the numerical expression of this ground state, we also have contributions from states such as $\ket{2000}$ but with a small coefficient. In general, for small $\beta$,
the two state approximation is good (the corrections are small).

A $n=3$ state becomes the ground state near $\mu$=3.03. The ground state can in good approximation be described as
\begin{equation}
\ket{\Omega, n=3}=\frac{1}{2}(\ket{0111 } +\ket{1011}+\ket{1101}+\ket{1110})\ ,
\end{equation}
which is $\ket{\Omega, n=1}$ with 0's and 1's interchanged and one can interpret the 0 as ``holes".
Finally,  near  $\mu$=3.10, $\ket{1111}$ becomes the ground states (with again many small corrections).
In general, there is approximate mirror symmetry about the ``half-filling" situation.

A similar approximate two state description is valid in the next SF phase located between the MI phases with $\rho$ 1 and 2.
One just needs to replace 0 by 1 and 1 by 2.

\begin{figure}[h]
  \includegraphics[width=0.2\textwidth,angle=0]{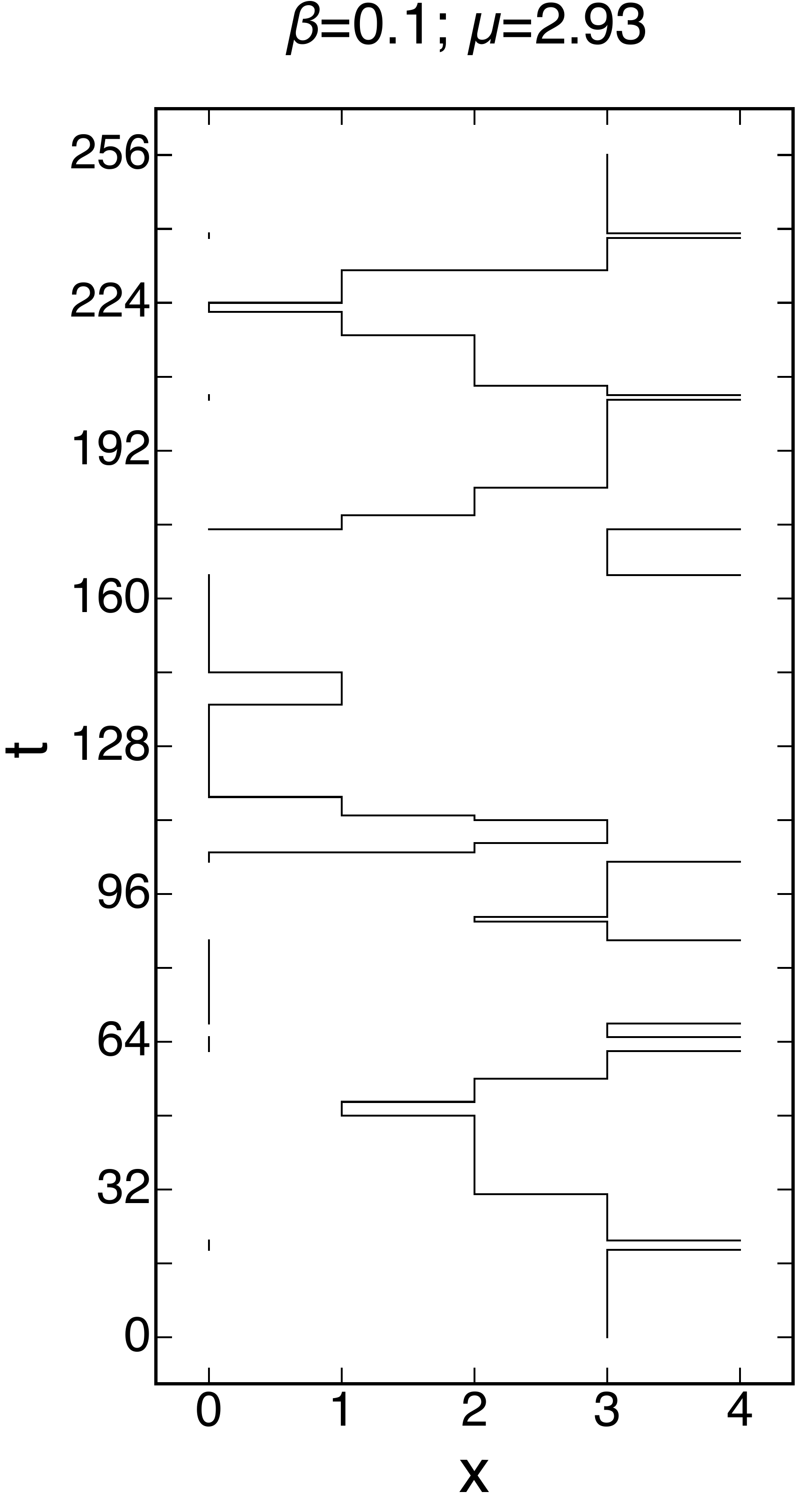}
   \includegraphics[width=0.2\textwidth,angle=0]{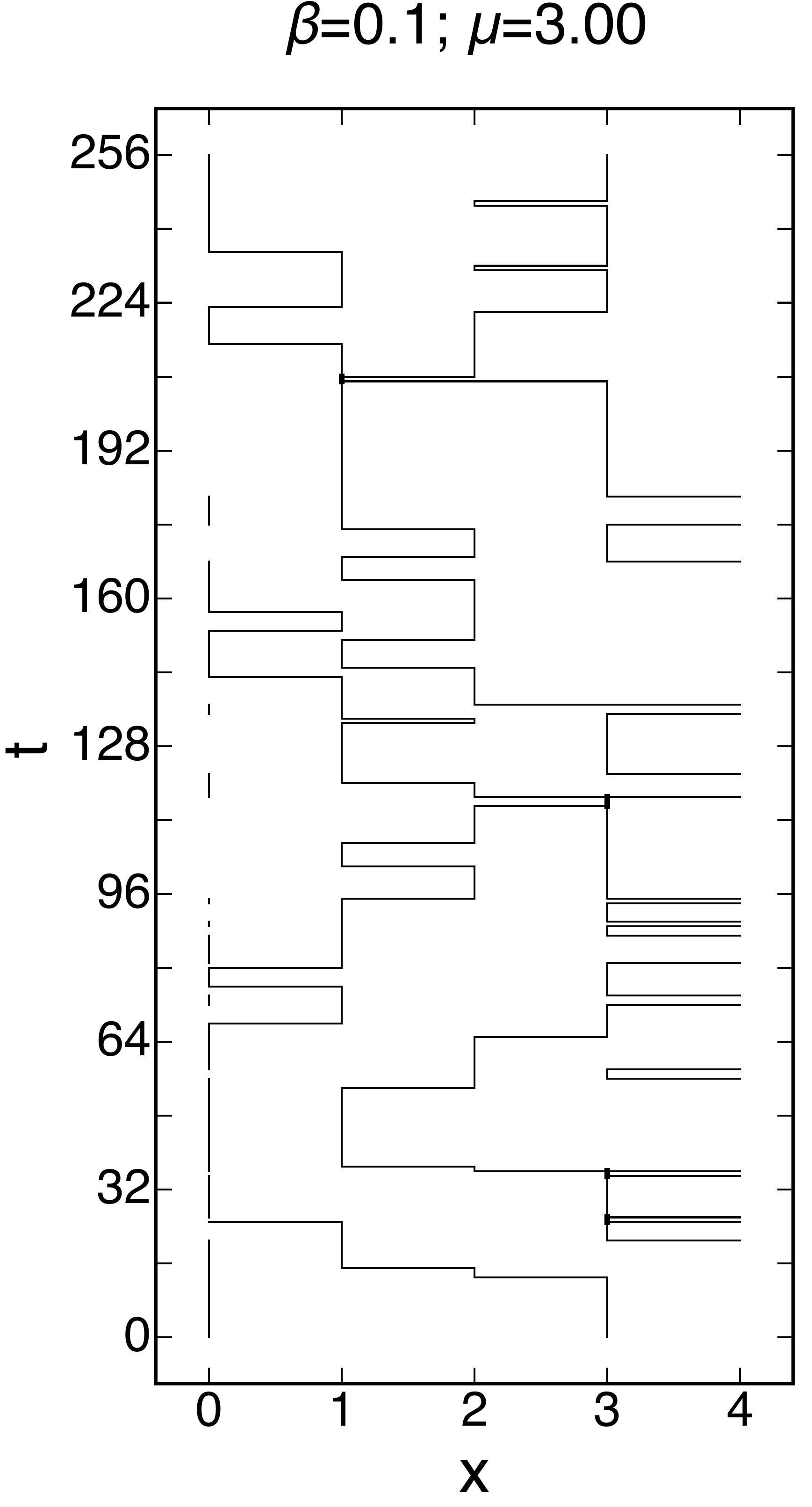}
   \includegraphics[width=0.2\textwidth,angle=0]{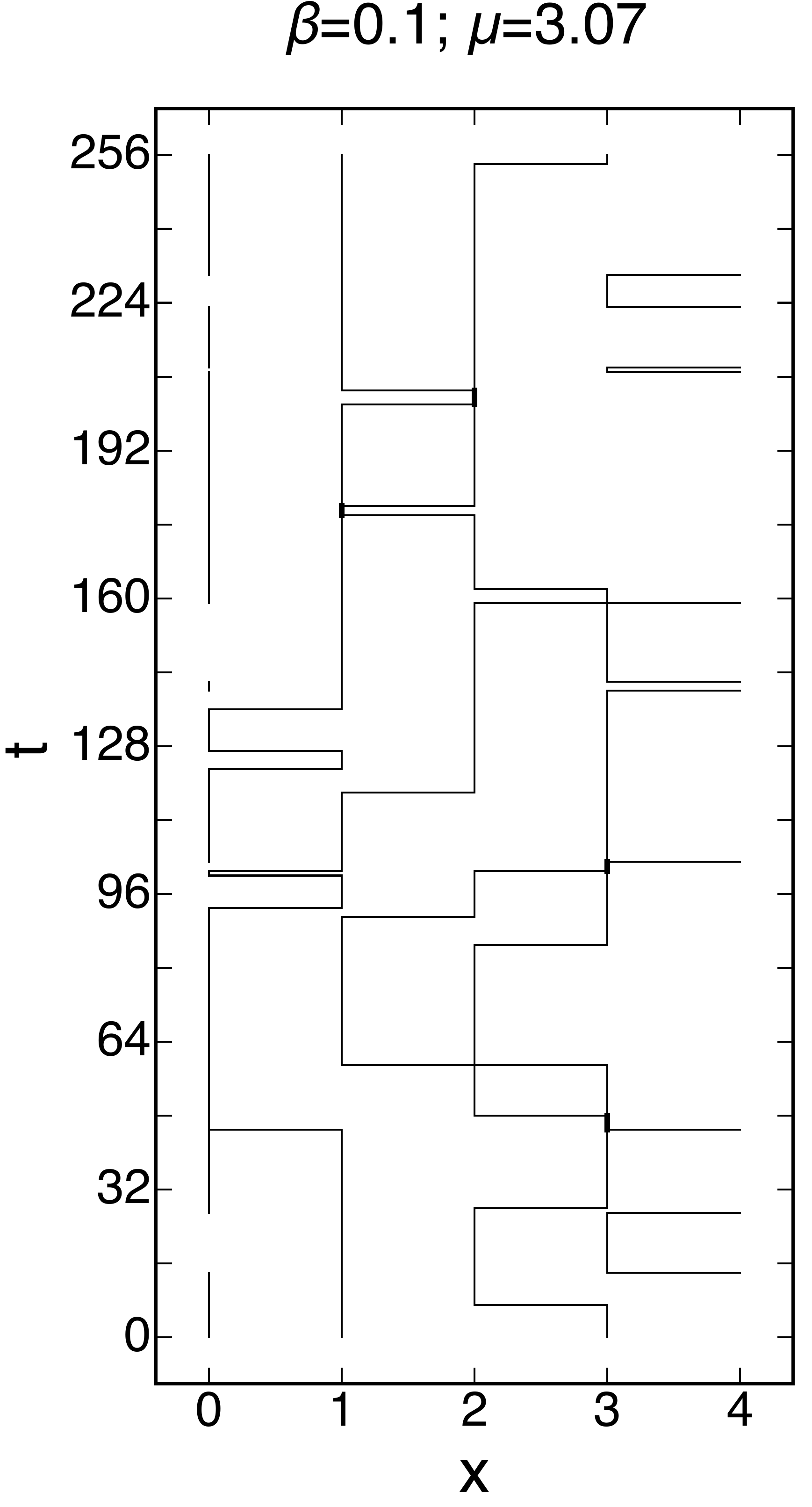}
   \includegraphics[width=0.2\textwidth,angle=0]{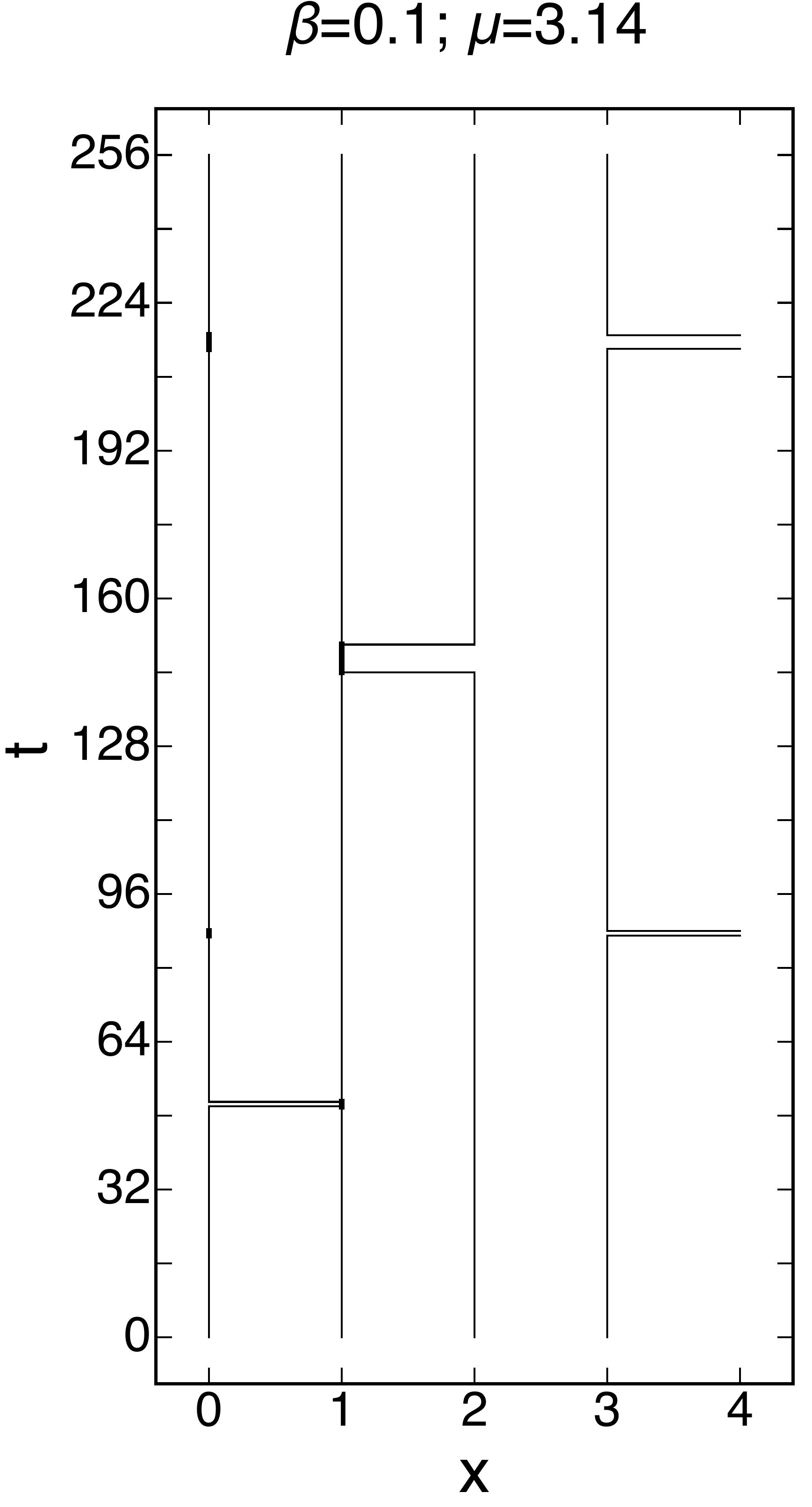}
\caption{ \label{fig:allworm} Worm configurations for  $\mu$ =2.93 ($n$=1), 3.00 ($n$=2), 3.07 ($n$=3) and 3.14 ($n$=4) for $\beta=0.1$, $L_x=4$ and $L_t=256$.}
\end{figure}

\begin{figure}
\centering
 \vskip-3.8cm
\subfigure{\label{fig:ETEN:a} 

  \includegraphics[width=3.5in]{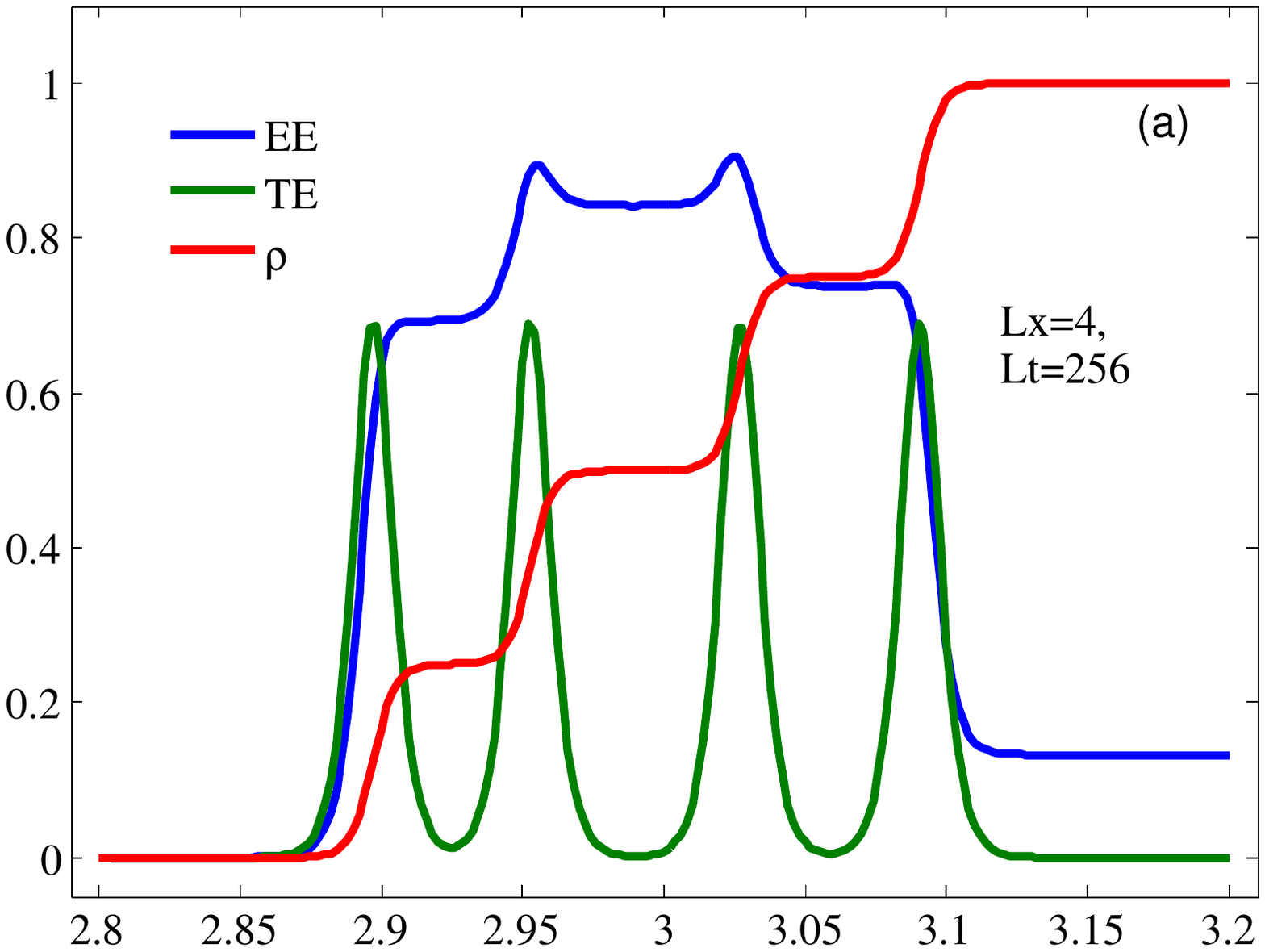}}
 \vskip-7.4cm
    \subfigure{\label{fig:ETEN:b} 
       \includegraphics[width=3.5in]{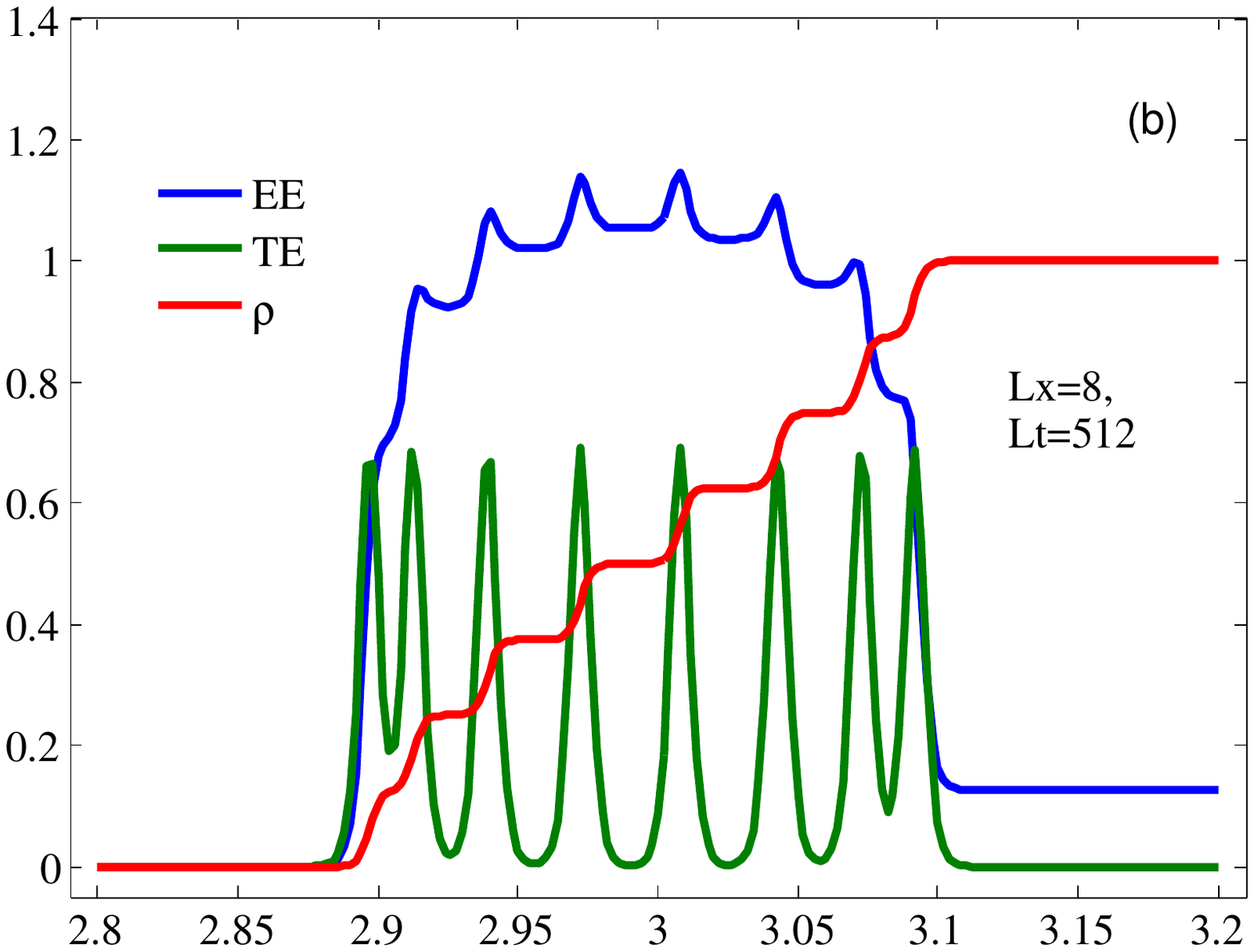}}
        \vskip-7.3cm
    \subfigure{\label{fig:ETEN:c} 
       \includegraphics[width=3.5in]{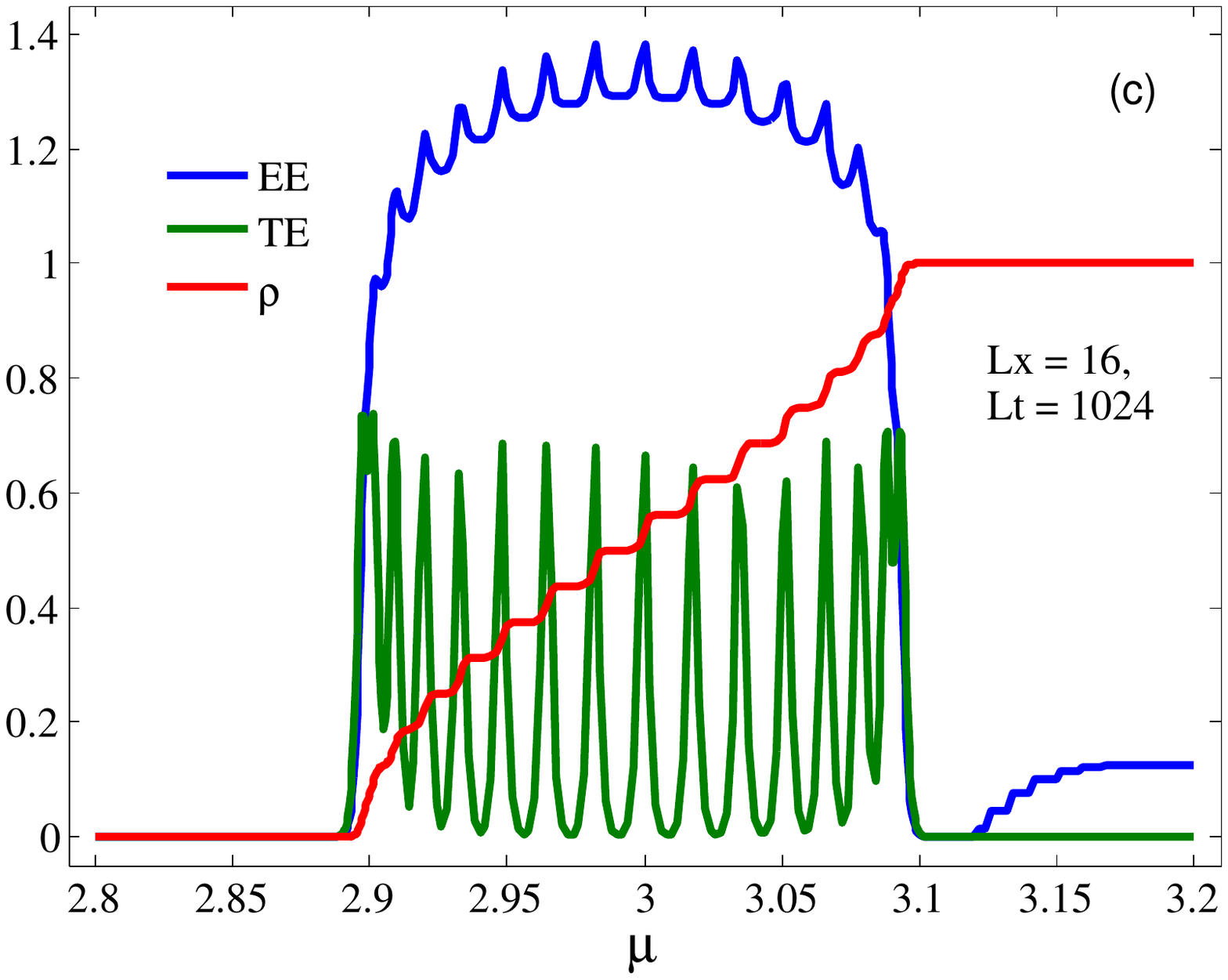}}
        \vskip-3.5cm
       \caption{ (Color online). Fine structure of the entanglement entropy (EE, blue), thermal entropy (TE, green) and particle density $\rho$ (red) spanning the MI($\rho=0$), SF and MI($\rho=1$) phases successively with $\mu$ increasing at fixed $\beta=0.1$. There are three different system sizes: $L_x=4$ with $L_t=256$, $L_x=8$ with $L_t=512$, and $L_x=16$  with $L_t=1024$, respectively. The thermal entropy has $L_x$ peaks culminating near $\ln 2 \simeq 0.69$; $\rho$ goes from 0 to 1 in $L_x$ steps and the entanglement entropy has an approximate mirror symmetry near half filling where it peaks.}
        \label{fig:ETEN} 
\end{figure}

We now follow the same path in the phase diagram but from the point of view of the word lines generated with the worm algorithm with $L_t$=256.
Typical world lines are displayed in  Fig. \ref{fig:typical}.
Given that $L_t$=256 is relatively large, by taking $\mu$ approximately in the middle of the density plateaus (discussed below, see Fig. \ref{fig:ETEN}), we obtain that most configurations have $n$ corresponding to the average value. Figure \ref{fig:allworm} shows typical results for  $\mu$ =2.93 ($n$=1), 3.00 ($n$=2), 3.07 ($n$=3) and 3.14 ($n$=4) using a graphical representation similar to Fig. \ref{fig:typical}.
Their spatial winding number are discussed in Appendix \ref{app:b}.
The most important feature of these configurations is that almost all the $|n|$' s are 0 or 1. For a given particle number $n$, between most time slices we have $n$ time links carrying a current 1 and $L_x-n$ time links carrying no current. In rare occasions, we observe that the line merge or cross. Overall, we can think of these configurations as a set of weakly interacting loops carrying a current 1.
This is in line with the dominance of the states like $\ket{1010}$ over states like $\ket{2000}$ discussed above.

We now proceed to calculate the \TE and the \EE for  $L_x=4$ with $L_t=256$ (as discussed above), and also for larger lattices $L_x=8$ with $L_t=512$, and $L_x=16$ with $L_t=1024$.
Figure \ref{fig:ETEN} shows the fine structure in the SF phase for increasing sizes. In each case, we go through the  MI($\rho=0$), SF, and MI($\rho=1$) phases successively as we keep increasing $\mu$ while keeping the
other parameters fixed, as already illustrated in Fig. \ref{fig:phased2s}. We kept the same truncation dimension $D_s=101$ for the three cases.

From  Fig.~\ref{fig:ETEN}, we observe an oscillation structure in the \EE and the thermal entropy.
The number of sites in the spatial direction $L_x$ dictates the
fine structure. There are $L_x$ transition points in the \EE and $L_x$ peaks in the thermal entropy. With $L_x$ increasing, the transition points close to the two boundaries become
difficult to resolve as shown in the case
of $L_x=16$, $L_t=1024$ in Fig. \ref{fig:ETEN:c}. Higher $\mu$ resolution and larger $D_s$ are needed to obtain a clear picture of the oscillations. The approximate mirror symmetry of the \EE with respect to the half-filling point persists for larger lattices. As far as the \TE is concerned, the height of the peaks are all close to $\ln 2$, which corresponds to the two-fold degeneracy in the ground state. The peaks are located at the place where the energy levels from the ground state and the first excited state cross as shown in Fig.~\ref{fig:EL}.

As $L_x$ increases, the fine structure near the two boundaries connected to the MI phases is not so prominent as the middle regime and
a better  resolution is needed to discern the fine structure. A similar discussion applies to the situation when the system experience the MI$(\rho=i)$, SF, MI($\rho=i+1$) phases successively and we have checked that the same type of fine structure appears.
Note also that for small $L_x$ and $L_t$ large enough, $\rho$ has significant plateaus that could be qualified as incompressible regions, however the width of these regions shrinks like $1/L_x$ as $L_x$ increases.
\section{Conclusions}
\label{sec:conclusions}
\label{sec:con}
\begin{figure}[h]
\vskip-3.8cm
\advance\leftskip 0.2cm
  \includegraphics[width=0.55\textwidth,angle=0]{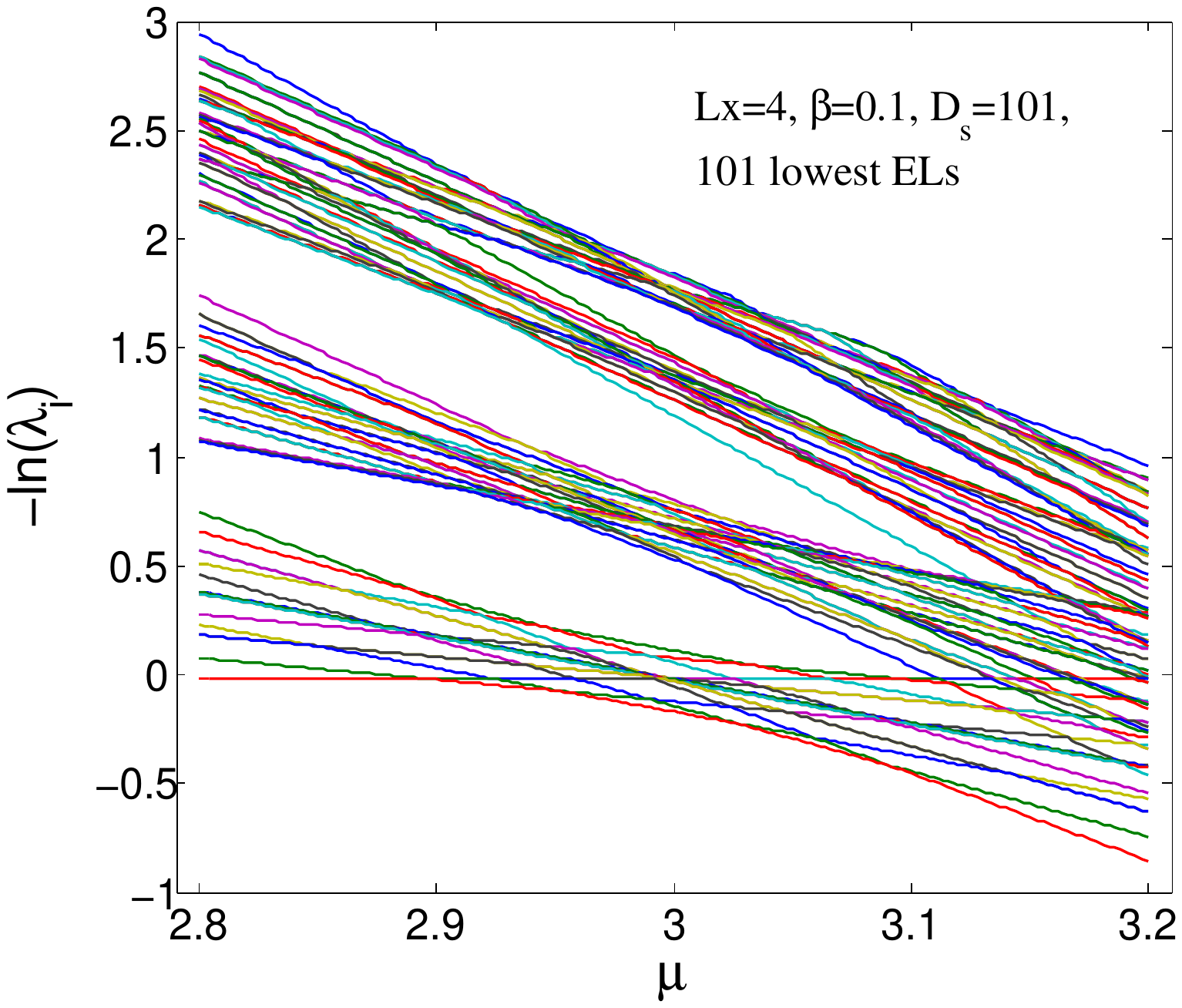}
\vskip-3.8cm
\caption{ \label{fig:Els}  (Color online). Energy levels for $\beta=0.1,L_x=4$.}
\end{figure}

In summary, we have used the TRG method combined with the conservation law to calculate the particle number density, the \TE and the \EE for the classical XY model with a chemical potential. The particle number distributions agree well but not perfectly with the results obtained with the worm algorithm. The discrepancies seem to increase with $L_x$.  This can probably be explained by
the fact that at small $\beta$, the spectrum has bands with $L_x$, $\frac{L_x(L_x-1)}{2}$ etc ... states  with increasing $n$.
Some of these bands merge when $\mu$ is increased enough to get in the SF phase. As $L_x$ increases, the bands become denser and we will typically keep the lowest energy states irrespectively of their particle number. Taking into account the particle number in the truncation process may help getting more accurate distributions. The energy bands are illustrated in Fig. \ref{fig:Els}.

Besides the large structure associated with the alternation between SF and MI phases with integer particle number density, we found that in the SF regime, there is an interesting fine structure controlled  by the spatial dimension $L_x$.
The entanglement entropy, the thermal entropy and the particle number density vary in a way that is consistent with each other.  The \TE shows $L_x$ peaks located at where the energy levels from the ground state and first excited state cross. Degenerate perturbation theory explains why the energy levels cross while varying the chemical potential.  As a result, the step-wise structure occurs in the particle number density in the SF regime, as already observed in Ref. \onlinecite{Banerjee:2010kc}. The particle number and the winding number of the world lines increase by one unit at a time. The \EE shows $L_x$ steps and an approximate mirror symmetry with respect to the half-filling point. The details of the fine structure depend on the ratio $L_x/L_t$
and the infinite volume limit of the two-dimensional classical model needs to be defined carefully.

In the approximation where the eigenstates of the transfer matrix are made of states with only 0 and 1 for all sites, or correspondingly if the world lines have mostly
links carrying currents with $|n|$ equal to 0 or 1, this symmetry corresponds to interchanging 0 and 1 and the particle number $n$ with $L_x-n$ which explains the approximate mirror symmetry of the entanglement entropy.
It is possible that the fermionic picture would become more clear if we could establish some approximate correspondence with the spin-1/2 quantum XY model.

We also would like to mention some analogies. As the chemical potential can be interpreted as an imaginary gauge field in the time direction, it is not surprising that studies of
Polyakov's loop \cite{Hands:2010zp,Hands:2010vw} (Wilson loops closing in the periodic time direction) show similar fine structure. Note also that the crossing pattern of the energy levels as a function of $\mu$ found here resembles the energy crossing found in a study of the spectrum of rotating tubes as a function of the rotation rate for a proposed cold atom experiments \cite{nateprogress}.

\appendix
\section{Remarks about Fig. \ref{fig:typical}}
\label{app:a}

In this appendix, we discuss the sign convention and the spatial winding number of Fig. \ref{fig:typical}. We also explain why this configuration is typical.

The sign of the $n$'s associated with links with $|n|$=1 can be figured out from the following information.  All the (vertical) temporal link indices are positive.
The sign of the (horizontal) spatial links can be obtained using the conservation law. On time slices 5, 7 and 17, the current moves to the right and the sign is (by convention) positive. On time slices 21 and 22, the current moves to the left and the sign is negative. In this configuration, the winding number in the temporal direction is 2 and the winding number in the spatial direction is 0 (there are as many positive as negative spatial $n$'s).

The fact that the configuration of Fig.  \ref{fig:typical} is typical for  $\beta$=0.1 and $\mu$=3 can be understood from the numerical values of the weights.
Because of the large value for $\mu$, the weight for the temporal links with $n$=0 ($I_0(0.1)\simeq 1.0025$) and $n$=1  ($I_1(0.1)\exp(3)\simeq 1.00553$) are almost the same, while the weight for $n$=2 ($I_2(0.1)\exp(6)\simeq 0.5047$) is smaller. The weight for $n$=-1 ($I_1(0.1)\exp(-3)\simeq 0.002492$) is very small and there are no
temporal links with negative values of $n$ in the configuration. For the spatial links, the relative cost of a lateral move is $I_1(0.1)/I_0(0.1) \simeq 0.05$ and there are only
6 lateral moves. The fact that there are only 2 temporal links with $n$=2 can be understood from the fact that the merging of two $n$=1 lines into one $n$=2 line
requires one lateral move in addition of a weight about twice smaller.

\section{Spatial winding numbers in Fig. \ref{fig:allworm}}
\label{app:b}
In this appendix, we discuss the spatial winding numbers of the configurations of Fig. \ref{fig:allworm}.
For  $\mu$=2.93, all the time (vertical)  links have $n$=1.
For the spatial links there are 20 right movers and 16 left movers (so the spatial winding number is 1).
For  $\mu$=3.00 ($n$=2), between most time slices, there are two vertical links carrying a $n$=1 current, and the two lines only merge 4 times into a single $n$=2 line for a small number of time steps (the total is
5 vertical links with $n$=2). There are 26 right movers and 34 left movers (so the spatial winding number is -2).
For  $\mu$=3.07 ($n$=3),
between most time slices, there are three vertical links carrying a $n$=1 current,
There are 5 occurrences where two $n$=1 merge into a single $n$=2 line for a small number of time steps (the total is
10 vertical links with $n$=2). There are also 3 crossing (points where the four lines attached all have $|n|$=1). There are 19 right movers and 15 left movers (so the spatial winding number is +1). For  $\mu$=3.14 ($n$=4), $\beta=0.1$, $L_x=4$ and $L_t=256$, there are essentially four parallel vertical lines each carrying a $n$=1 current except for 4 occurrences where they briefly merge (11 $n$=2 vertical lines, 4 right movers, 4 left movers, no spatial winding number).

\begin{acknowledgments}
We thank S. Chandrasekharan for help with the worm algorithm code and discussions of the fermionic picture, J. Unmuth-Yockey and J. Osborn for discussions about TRG calculations, M. C. Ba\~nuls for discussions about the entanglement entropy and T. Xiang for insightful discussions.
This work started during the workshop Precision Many-Body Physics of Strongly Correlated Quantum Matter at the KITPC in Beijing in May 2014 and the manuscript was being written while Y. M. attended the workshop Understanding Strongly Coupled Systems in High
Energy and Condensed Matter Physics at the Aspen Center for Physics in May-June 2015.
We thank the organizers and participants of these two workshops for many stimulating discussions. This research was supported in part  by the Department of Energy
under Award Number DOE grant DE-SC0010114, and by the Army Research Office of the Department of Defense under Award Number W911NF-13-1-0119.
This work utilized the Janus supercomputer, which is supported by the National Science Foundation (award number CNS-0821794) and the University of Colorado Boulder. The Janus supercomputer is a joint effort of the University of Colorado Boulder, the University of Colorado Denver and the National Center for Atmospheric Research.
The work was supported by Natural Science Foundation of China for the Youth (Grants No.11304404).
L.-P.Yang thanks Zi-Xiang Hu's computational resources
and Hai-Qing Lin 's cordial invitation for visiting CSRC.
\end{acknowledgments}
\bibliography{../../centraltex/centralmacbib.bib}


\end{document}